\documentclass[prd,floatfix,nofootinbib,eqsecnum,showpacs]{revtex4}
\usepackage{epsfig}
\usepackage{amssymb,amsmath}
\newcommand{\eq}[1]{Eq.~(\ref{#1})}

\newcommand{\be}{\begin{equation}}
\newcommand{\ee}{\end{equation}}
\newcommand{\bea}{\begin{eqnarray}}
\newcommand{\eea}{\end{eqnarray}}
\newcommand{\ben}{\begin{eqnarray*}}
\newcommand{\een}{\end{eqnarray*}}

\newcommand{\DS}{Dyson-Schwinger }

\newcommand{\w}{\omega}
\newcommand{\e}{\varepsilon}
\newcommand{\al}{\alpha}
\newcommand{\ba}{\beta}
\newcommand{\ga}{\gamma}
\newcommand{\G}{\Gamma}
\newcommand{\de}{\delta}

\newcommand{\si}{\sigma}

\newcommand{\ro}{\rho}
\newcommand{\la}{\lambda}
\newcommand{\La}{\Lambda}
\newcommand{\ka}{\kappa}
\newcommand{\ta}{\tau}

\newcommand{\ha}{\frac{1}{2}}
\newcommand{\pd}{\partial}

\renewcommand{\th}{\theta}

\newcommand{\cd}{{\cal D}}
\newcommand{\cs}{{\cal S}}
\newcommand{\cc}{{\cal C}}
\newcommand{\co}{{\cal O}}
\renewcommand{\div}{\vec{\nabla}}

\newcommand{\s}[2]{{#1}\!\cdot\!{#2}}
\newcommand{\ov}[1]{\overline{#1}}
\newcommand{\dk}[1]{\,\,\,\raisebox{-0.4ex}{\large $\bar{}$}\!\!d\,{#1}\,}

\newcommand{\ev}[1]{<\!\!{#1}\!\!>}

\begin{document}
\title{Leading order infrared quantum chromodynamics in Coulomb gauge}
\author{P.~Watson}
\author{H.~Reinhardt}
\affiliation{Institut f\"ur Theoretische Physik, Universit\"at T\"ubingen, 
Auf der Morgenstelle 14, D-72076 T\"ubingen, Deutschland}
\begin{abstract}
A truncation scheme for the Dyson-Schwinger equations of quantum 
chromodynamics in Coulomb gauge within the first order formalism is 
presented.  The truncation is based on an Ansatz for the Coulomb kernel 
occurring in the action.  Results at leading loop order and in the infrared 
are discussed for both the Yang-Mills and quark sectors.  It is found that 
the resulting equations for the static gluon and quark propagators agree 
with those derived in a quasi-particle approximation to the canonical 
Hamiltonian approach.  Moreover, a connection to the heavy quark limit is 
established.  The equations are analyzed numerically and it is seen that in 
both the gluonic and quark sectors, a nontrivial dynamical infrared mass 
scale emerges.
\end{abstract}
\maketitle
\section{Introduction}
\setcounter{equation}{0}

Confinement and the dynamical breaking of chiral symmetry are two highly 
nontrivial, nonperturbative aspects of the hadron spectrum.  Ideally, we 
would like to understand both from first principles calculations of the 
underlying gauge theory which is quantum chromodynamics (QCD).  Because both 
effects are manifested in the infrared regime (and where singularities may 
occur), nonperturbative continuum functional methods are one framework 
within which to study the problem.  In such studies, one is invariably 
forced to choose a gauge and our choice is Coulomb gauge.  One initial 
reason for this choice is that there exists an appealing picture for 
confinement: the Gribov-Zwanziger scenario 
\cite{Gribov:1977wm,Zwanziger:1995cv,Zwanziger:1998ez}.

Coulomb gauge studies of QCD have become an area of increasing importance in 
recent years, mainly due to progress within the canonical Hamiltonian 
approach (see, for example, Refs.~\cite{Szczepaniak:1995cw,
LlanesEstrada:2000jw,Szczepaniak:2001rg,Szczepaniak:2003ve,Reinhardt:2004mm,
Feuchter:2004mk,Campagnari:2010wc,Pak:2011wu} and references therein), 
although such studies have been around for quite some time 
(e.g., Refs.~\cite{Adler:1984ri,Schutte:1985sd}).  The essential idea is 
that given the Hamilton operator \cite{Christ:1980ku} and an Ansatz for the 
ground state vacuum wavefunctional (supplemented with Gauss' law), the 
variational principle can then be used to generate equations for the various 
Green's functions of the theory.  For a given Ansatz, the resulting 
equations are exact.

A second approach to nonperturbative QCD in Coulomb gauge (and that 
considered here) is to study the \DS equations \cite{Watson:2006yq}.  In 
contrast to the canonical approach, \DS studies in Coulomb gauge are far 
less developed mainly due to their inherent technical challenges (although 
attempts have been made \cite{Lichtenegger:2009dw,Alkofer:2009dm}).  This 
being said, various facets of the formalism have been understood: their 
one-loop perturbative behavior 
\cite{Watson:2007mz,Watson:2007vc,Popovici:2008ty}, the Slavnov-Taylor 
identities \cite{Watson:2008fb}, the emergence of a nonperturbative 
constraint on the total charge \cite{Reinhardt:2008pr} and the case of 
heavy quarks \cite{Popovici:2010mb,Popovici:2010ph,Popovici:2011yz}.  The 
standard \DS approach is distinct from the canonical approach in two 
primary respects: it includes the energy dependence of the Green's 
functions from the outset and must therefore deal directly with the 
noncovariance of Coulomb gauge; also, truncations are typically made to the 
individual terms in the (full) equations.

Clearly, it is desirable to be able to connect the canonical Hamiltonian 
and \DS formalisms.  Working in concert, the two different formulations can 
mutually reinforce each other -- what is difficult in one may be intuitive 
in the other and vice versa.  An initial aim of this paper is thus to show 
how the gap equations for the static gluon and quark propagators, obtained 
originally within the canonical formalism 
\cite{Szczepaniak:2001rg,Adler:1984ri}, can be derived from a leading order 
truncation of the \DS equations.  This truncation treats quarks and gluons 
on an equal footing.  It also takes into account the nonperturbative 
constraint that the total color charge be conserved and vanishing 
\cite{Reinhardt:2008pr}.  Moreover, the connection to the known Coulomb 
gauge heavy quark limit 
\cite{Popovici:2010mb,Popovici:2010ph,Popovici:2011yz} can be established.  
The gap equations will be analyzed numerically for a particular 
nonperturbative input and various aspects will be explored.  In the case of 
the gluon sector, it will be seen that a nontrivial dynamical mass scale 
emerges and which is connected to the nonperturbative renormalization of 
the theory.  For the quark sector, both the dynamical chiral symmetry 
breaking and the heavy quark limit will be discussed.

The paper is organized as follows.  We begin in Sec.~\ref{sec:first} by 
introducing the first order formalism, including the total color charge 
constraint that arises from the resolution of the temporal zero modes 
inherent to Coulomb gauge.  The \DS equations and the leading order 
truncation scheme will be presented in Sec.~\ref{sec:dses}.  The analytic 
development of the \DS equations, in particular the reduction to the static 
equations and the heavy quark limit, is given in Sec.~\ref{sec:andev}.  
Numerical results for both the gluonic and quark sectors appear in 
Sec.~\ref{sec:numres}.  We close the paper with a summary and discussion.

\section{\label{sec:first}First order formalism}
\setcounter{equation}{0}
To begin, let us consider the generating functional integral and action for 
QCD and review various aspects of the first order formalism in Coulomb 
gauge.  The description initially follows closely in the spirit of 
Refs.~\cite{Zwanziger:1998ez,Cucchieri:2000hv,Watson:2006yq,
Reinhardt:2008pr}.  The generating functional is written as
\be
Z\left[\ro,\vec{J},\ov{\chi},\chi\right]
=\int\cd\Phi\exp{\left\{\imath\cs_{QCD}
+\imath\int dx\left[\ro_x^a\si_x^a+\s{\vec{J}_x^a}{\vec{A}_x^a}
+\ov{\chi}_{\al x}q_{\al x}+\ov{q}_{\al x}\chi_{\al x}\right]\right\}}
\ee
where $\cd\Phi$ generically represents the functional integral measure over 
all fields present, with sources $\ro$, $\vec{J}$, $\ov{\chi}$, $\chi$ for 
the various gluon and quark fields (see below).  In this section, only the 
source $\ro$ will be relevant to the discussion and so we set 
$\vec{J}=\ov{\chi}=\chi=0$.  The (Minkowski space) QCD action reads
\be
\cs_{QCD}=\int dx\left\{\ov{q}_{\al x}\left[\imath\ga^0D_{0x}
+\imath\s{\vec{\ga}}{\vec{D}_{x}}-m\right]_{\al\ba}q_{\ba x}
+\frac{1}{2}\s{\vec{E}_x^a}{\vec{E}_x^a}
-\frac{1}{2}\s{\vec{B}_x^a}{\vec{B}_x^a}\right\}.
\ee
In the above, ($\ov{q}$) $q_{\ba x}$ represents the (conjugate) quark field 
with fundamental color, spin and flavor indices collectively denoted with 
the index $\ba$ and position argument denoted with subscript $x$.  The 
Dirac $\ga$-matrices obey the Clifford algebra 
$\{\ga^\mu,\ga^\nu\}=2g^{\mu\nu}$ with metric 
$g^{\mu\nu}=\mbox{diag}(1,-\vec{1})$ (we explicitly extract all minus signs 
associated with the metric such that the components of spatial vectors such 
as $\vec{x}$ are written with subscripts, i.e., $x_i$).  The temporal and 
spatial components of the covariant derivative in the fundamental color 
representation are
\bea
D_{0x}&=&\pd_{0x}-\imath g\si_x=\pd_{0x}-\imath g\si_x^aT^a,\nonumber\\
\vec{D}_x&=&\div_x+\imath g\vec{A}_x=\div_x+\imath g\vec{A}_x^aT^a
\eea
where $\si_x^a$ and $\vec{A}_x^a$ are the temporal and spatial components of 
the gluon field, respectively, and where the superscript $a$ denotes the 
color index in the adjoint representation.  The generators $T^a$ obey 
$[T^a,T^b]=\imath f^{abc}T^c$, where the $f^{abc}$ are the structure 
constants and we use the normalization $\mbox{Tr}[T^aT^b]=\de^{ab}/2$.  The 
chromoelectric and chromomagnetic fields are written in terms of the gluon 
field as
\bea
\vec{E}_x^a&=&-\pd_{0x}\vec{A}_x^a-\vec{D}_x^{ab}\si_x^b,\nonumber\\
\vec{B}_x^a&=&\div_x\times\vec{A}_x^a
-\frac{1}{2}gf^{abc}\vec{A}_x^b\times\vec{A}_x^c
\eea
with the spatial component of the covariant derivative in the adjoint color 
representation given by
\be
\vec{D}_x^{ab}=\de^{ab}\div_x-gf^{acb}\vec{A}_x^c.
\label{eq:covadj0}
\ee

The QCD action is invariant under gauge transforms of the type
\bea
\si&\rightarrow&\si^\th
=U\si U^\dag-\frac{\imath}{g}\left(\pd_0U\right)U^\dag,\nonumber\\
\vec{A}&\rightarrow&\vec{A}^\th
=U\vec{A}U^\dag+\frac{\imath}{g}\left(\div U\right)U^\dag,\nonumber\\
q&\rightarrow&q^\th=Uq
\eea
where $U_x=\exp{\{-\imath\th_x^aT^a\}}$ is a spacetime dependent element of 
the $SU(N_c)$ group.  Because of this invariance, the functional integral 
(in the absence of sources) contains a divergence by virtue of the 
integration over the gauge group.  Whilst this is in principle a global 
factor which can be absorbed into the normalization, when calculating 
Green's functions it leads to the necessity for fixing the gauge.  This is 
typically achieved in the continuum formalism via the Faddeev-Popov 
technique which involves inserting the following identity into the 
functional integral:
\be
\openone=\int\cd\th\de\left(F\left[\si^\th,\vec{A}^\th\right]\right)
\mbox{Det}\left[M^{ab}(x,y)\right],\;\;\;\;
M^{ab}(x,y)
=\left.\frac{\de F^a\left[\si_x^\th,\vec{A}_x^\th\right]}{\de\th_y^b}
\right|_{F=0}.
\label{eq:fp0}
\ee
The expression $F[\si,\vec{A}]$ determines the gauge condition ($F=0$) and 
we choose Coulomb gauge:
\be
F[\si,\vec{A}]:=\s{\div}{\vec{A}},
\ee
for which the Faddeev-Popov kernel reads
\be
M^{ab}(x,y)\sim-\s{\div}{\vec{D}_x^{ab}}\de(x-y).
\ee
There are caveats to the identity \eq{eq:fp0}, namely that when the gauge 
fixing is incomplete, zero modes of the Faddeev-Popov operator will arise 
and one encounters the Gribov problem \cite{Gribov:1977wm}.  In Coulomb 
gauge there is a special case: temporal zero modes corresponding to time 
dependent but spatially independent gauge transforms 
\cite{Reinhardt:2008pr}, arising because the Faddeev-Popov operator involves 
only spatial differential operators, and for which
\be
-\s{\div}{\vec{D}_x^{ab}}\th^b(x_0)=0.
\ee
Clearly, such zero eigenvalues (there are $N_c^2-1$ of them at each time 
$x_0$) for the Faddeev-Popov operator automatically result in a vanishing 
functional determinant and hence invalidate the identity \eq{eq:fp0}.  In 
this work, we shall not consider the more general case of spatially 
dependent zero modes (which lead to the existence of Gribov copies in the 
usual sense).  Following \cite{Reinhardt:2008pr}, we modify the original 
identity to
\be
\openone=\int\cd\ov{\th}\de\left(F\left[\si^\th,\vec{A}^\th\right]\right)
\ov{\mbox{Det}}\left[M^{ab}(x,y)\right]
\label{eq:fp1}
\ee
where $\cd\ov{\th}$ explicitly excludes the temporal zero modes $\th(x_0)$ and
\be
\ov{\mbox{Det}}\left[M^{ab}(x,y)\right]
=\mbox{Det}\left[M^{ab}(x,y)\right]_{-\s{\div}{\vec{D}}\th\neq0}
\ee
is the determinant with such zero modes removed.  Replacing the source term 
($\ro$) for the temporal gluon field in the functional integral 
(as mentioned, the other sources play no role in this section and are set 
to zero), the generating functional for Coulomb gauge QCD is thus written
\be
Z\left[\ro\right]=
\int\cd\Phi\de\left(\s{\div}{\vec{A}}\right)
\ov{\mbox{Det}}\left[-\s{\div}{\vec{D}}\right]
\exp{\left\{\imath\cs_{QCD}+\imath\int dx\ro_x^a\si_x^a\right\}}.
\ee

To proceed, it is useful to convert to the first order formalism 
\cite{Zwanziger:1998ez,Watson:2006yq,Reinhardt:2008pr}.  This is achieved by 
introducing an auxiliary vector field ($\vec{\pi}$), noting the following 
functional integral identity for the chromoelectric part of the action:
\be
\exp{\left\{\imath\int dx\frac{1}{2}\s{\vec{E}_x^a}{\vec{E}_x^a}\right\}}
=\int\cd\vec{\pi}\exp{\left\{\imath\int dx\left[
-\frac{1}{2}\s{\vec{\pi}_x^a}{\vec{\pi}_x^a}
-\s{\vec{\pi}_x^a}{\vec{E}_x^a}\right]\right\}}.
\ee
The new field is then split into components using
\be
\mbox{const}=\int\cd\phi\cd\ta\exp{\left\{-\imath\int dx\,\ta_x^a
\left(\s{\div_x}{\vec{\pi}_x^a}+\div_x^2\phi_x^a\right)\right\}},
\ee
changing variables $\vec{\pi}\rightarrow\vec{\pi}-\div\phi$ and integrating 
out the Lagrange multiplier.  The generating functional now reads
\be
Z\left[\ro\right]=
\int\cd\Phi\de\left(\s{\div}{\vec{A}}\right)
\de\left(\s{\div}{\vec{\pi}}\right)
\ov{\mbox{Det}}\left[-\s{\div}{\vec{D}}\right]
e^{\imath\cs}
\ee
with the action
\bea
\cs&=&\cs_q+\cs'+\cs_\si,\nonumber\\
\cs_q&=&\int dx\,\ov{q}_{\al x}\left[\imath\ga^0\pd_{0x}
+\imath\s{\vec{\ga}}{\vec{D}}-m\right]_{\al\ba}q_{\ba x},\nonumber\\
\cs'&=&\int dx\left[-\ha\s{\vec{B}_x^a}{\vec{B}_x^a}
-\ha\s{\vec{\pi}_x^a}{\vec{\pi}_x^a}+\s{\vec{\pi}_x^a}{\pd_{0x}\vec{A}_x^a}
+\ha\phi_x^a\div_x^2\phi_x^a\right],\nonumber\\
\cs_{\si}&=&\int dx\,\si_x^a\left(\s{\div_x}{\vec{D}_x^{ab}}\phi_x^b
+\ro_x^a+gf^{abc}\s{\vec{A}_x^b}{\vec{\pi}_x^c}
+g\ov{q}_{\al x}[\ga^0T^a]_{\al\ba}q_{\ba x}\right).
\eea
Were the $\de$-functional constraints written in terms of Lagrange 
multiplier fields, the determinant written in terms of ghosts and sources 
for all fields present, the theory would be in a local form suitable for 
discussing perturbation theory as in 
Refs.~\cite{Watson:2006yq,Watson:2007mz,Popovici:2008ty}.  The important 
feature about the decomposition to the first order formalism is that the 
action is now linear in $\si$, which can be integrated out to give
\be
Z\left[\ro\right]=
\int\cd\Phi\de\left(\s{\div}{\vec{A}}\right)
\de\left(\s{\div}{\vec{\pi}}\right)
\ov{\mbox{Det}}\left[-\s{\div}{\vec{D}}\right]
\de\left(\s{\div}{\vec{D}}\phi+\ov{\ro}\right)
e^{\left(\imath\cs_q+\imath\cs'\right)}
\ee
where
\be
\ov{\ro}_x^a=\ro_x^a+gf^{abc}\s{\vec{A}_x^b}{\vec{\pi}_x^c}
+g\ov{q}_{\al x}[\ga^0T^a]_{\al\ba}q_{\ba x}.
\ee
The $\phi$ field can be integrated out by using the eigenfunctions of the 
Faddeev-Popov operator as a complete orthonormal basis for an expansion, 
the crucial point being that one must take into account the existence of 
the temporal zero modes.  Following Ref.~\cite{Reinhardt:2008pr}, the 
result is
\be
\int\cd\phi\de\left(\s{\div}{\vec{D}}\phi+\ov{\ro}\right)
\exp{\left\{\imath\int dx\,\ha\phi_x^a\div_x^2\phi_x^a\right\}}
=\de\left(\int d\vec{x}\,\ov{\ro}\right)
\ov{\mbox{Det}}\left[-\s{\div}{\vec{D}}\right]^{-1}
\exp{\left\{-\imath\int dx\,\ha\ov{\ro}_x^a\hat{F}_x^{ab}\ov{\ro}_x^b\right\}}
\ee
where
\be
\hat{F}_x^{ab}=\left[-\s{\div_x}{\vec{D}_x^{ac}}\right]^{-1}
\left(-\div_x^2\right)\left[-\s{\div_x}{\vec{D}_x^{cb}}\right]^{-1}.
\ee
The $\de$-functional constraint that emerges constrains the total color 
charge, the spatial integral arising from the projection onto the temporal 
zero mode.  The generating functional is now,
\be
Z[\ro]=\int\cd\Phi\de\left(\s{\div}{\vec{A}}\right)
\de\left(\s{\div}{\vec{\pi}}\right)\de\left(\int d\vec{x}\,\ov{\ro}\right)
e^{\left(\imath\cs_q+\imath\cs\right)},
\label{eq:genfunc0}
\ee
with the action term
\be
\cs=\int dx\left[-\ha\s{\vec{B}_x^a}{\vec{B}_x^a}
-\ha\s{\vec{\pi}_x^a}{\vec{\pi}_x^a}+\s{\vec{\pi}_x^a}{\pd_{0x}\vec{A}_x^a}
-\ha\ov{\ro}_x^a\hat{F}_x^{ab}\ov{\ro}_x^b\right].
\ee

Some comments are in order.  The (modified) Faddeev-Popov determinant in the 
generating functional cancels exactly, which is equivalent to saying that 
Coulomb gauge is ghost-free and this arises from the elimination of the 
temporal gluon field.  Notice though that the inverse Faddeev-Popov operator 
still plays a role.  What is left of the gluon sector concerns the two 
spatially transverse vector fields $\vec{A}$ and $\vec{\pi}$ ($\vec{\pi}$ 
would classically correspond to the momentum conjugate of $\vec{A}$) which 
in quantum electrodynamics would give rise to the two physical transverse 
polarization states of the photon.  The remnant $\de$-functional constraint 
is the statement that the total color charge of the system must be conserved 
and vanishing and this explicitly includes a contribution from the external 
source $\ro$, the gluon field and the quark term.  This is nothing other 
than the application of Gauss' law.

In order to make sense of the generating functional, \eq{eq:genfunc0}, the 
$\de$-functional constraint on the total charge must be rewritten in a 
useful form.  This is most conveniently done with a Gaussian form:
\be
\de\left(\int d\vec{x}\,\ov{\ro}\right)
\sim\lim_{\cc\rightarrow\infty}{\cal N}(\cc)
\exp{\left\{-\frac{\imath}{2}\int dy\,dz\,
\ov{\ro}^a(y)\cc\de^{ab}\de(y_0-z_0)\ov{\ro}^b(z)\right\}}
\ee
where $\cc$ is a constant, ${\cal N}(\cc)$ is a normalization factor (that 
will be henceforth included implicitly in the functional integral measure) 
and the limit $\cc\rightarrow\infty$ will be taken only at the end of any 
calculation.  The generating functional can thus be written in the form
\be
Z[\ro]\sim\lim_{\cc\rightarrow\infty}
\int\cd\Phi\de\left(\s{\div}{\vec{A}}\right)
\de\left(\s{\div}{\vec{\pi}}\right)
\exp{\left\{\imath\cs''\right\}}
\exp{\left\{-\frac{\imath}{2}\int dy\,dz\,\ov{\ro}^a(y)
\left[F^{ab}\left(y_0;\vec{y},\vec{z}\right)+\cc\de^{ab}\right]
\de(y_0-z_0)\ov{\ro}^b(z)\right\}}
\label{eq:genfunc1}
\ee
where 
\be
\cs''=\cs_q+\int dx\left[-\ha\s{\vec{B}_x^a}{\vec{B}_x^a}
-\ha\s{\vec{\pi}_x^a}{\vec{\pi}_x^a}+\s{\vec{\pi}_x^a}{\pd_{0x}\vec{A}_x^a}
\right]
\ee
and
\be
F^{ab}\left(y_0;\vec{y},\vec{z}\right)
=\hat{F}_y^{ab}\de\left(\vec{y}-\vec{z}\right).
\ee
We shall refer to $F$ as the Coulomb kernel.  $F$ is defined at each time, 
$y_0$, since it involves no temporal operators.   We see that the effect of 
the total charge conservation constraint is the presence of a potentially 
ill-defined and divergent constant additive term to the Coulomb kernel.  It 
will be seen later that the constant, $\cc$, cancels out when considering 
physical quantities.

As noted in Ref.~\cite{Cucchieri:2000hv}, the Coulomb kernel is intimately 
related to the temporal gluon propagator.  Although the temporal gluon field 
has been integrated out, the propagator still exists as a functional second 
derivative of $Z[\ro]$ (this is why we have so far retained the source) and 
is defined in configuration space as
\be
W_{\si\si}^{cd}(v,w)=\left.\frac{1}{Z[\ro]}
\frac{\de^2Z[\ro]}{\de\imath\ro_v^c\de\imath\ro_w^d}\right|_{\ro=0}.
\label{eq:tempg0}
\ee
The temporal gluon propagator is known perturbatively in both the first and 
second order Coulomb gauge formalisms \cite{Watson:2007mz,Watson:2007vc}.  
Indeed, since the manipulations involved in going from the second to first 
order formalisms are integral identities performed on the generating 
functional, the spatial and temporal gluon propagators are identical in both 
formalisms.  Taking two functional derivatives of $Z$ given by 
\eq{eq:genfunc1} with respect to the source $\ro$ and subsequently setting 
$\ro=0$, one obtains:
\bea
\lefteqn{
W_{\si\si}^{cd}(v,w)
=\frac{1}{Z[0]}\int\cd\Phi\de\left(\s{\div}{\vec{A}}\right)
\de\left(\s{\div}{\vec{\pi}}\right)\exp{\left\{\imath\cs''\right\}}
}&&\nonumber\\
&&\!\!\times\lim_{\cc\rightarrow\infty}
\exp{\left\{-\frac{\imath}{2}\int dy\,dz\,
\hat{\ro}^a(y)\left[F^{ab}\left(y_0;\vec{y},\vec{z}\right)+\cc\de^{ab}\right]
\de(y_0-z_0)\hat{\ro}^b(z)\right\}}\nonumber\\
&&\!\!\times\left\{\imath\left[F^{cd}(v_0;\vec{v},\vec{w})+\cc\de^{cd}\right]
\de(v_0-w_0)
+\int d\vec{y}\,\left[F^{ce}(v_0;\vec{v},\vec{y})+\cc\de^{ce}\right]
\hat{\ro}^e(v_0,\vec{y}) \int d\vec{z}\,\left[F^{df}(w_0;\vec{w},\vec{z})
+\cc\de^{df}\right]\hat{\ro}^f(w_0,\vec{z})\right\}\nonumber\\
\label{eq:tempg1}
\eea
where
\be
\hat{\ro}_x^a=gf^{abc}\s{\vec{A}_x^b}{\vec{\pi}_x^c}
+g\ov{q}_{\al x}[\ga^0T^a]_{\al\ba}q_{\ba x}.
\ee
Were it not for the $\de$-functional constraint on the total charge, this 
expression would be the same as in Ref.~\cite{Cucchieri:2000hv} and the 
conclusion would be that the temporal gluon propagator splits into two 
parts, instantaneous and non-instantaneous:
\be
W_{\si\si}^{cd}(v,w)\sim\ev{\imath F^{cd}(v_0;\vec{v},\vec{w})}\de(v_0-w_0)
+\ev{\left[F_v^{ce}\hat{\ro}^e(v_0,\vec{v})\right]
\left[F_w^{df}\hat{\ro}^f(w_0,\vec{w})\right]},
\ee
the instantaneous part being the expectation value of the Coulomb kernel 
and arising because the original Faddeev-Popov operator involves only 
spatial operators.  However, the presence of the total charge constraint 
alters this: in particular, noting that the additional term is independent 
of the fields, we see that the instantaneous part of the temporal gluon 
propagator has the form
\be
W_{\si\si}^{cd}(v,w)^{\mbox{inst}}
\sim\ev{\imath F^{cd}(v_0;\vec{v},\vec{w})}\de(v_0-w_0)
+\lim_{\cc\rightarrow\infty}\imath\cc\de^{cd}\de(v_0-w_0)
\ee
(in the original term involving the expectation value of $F$, the 
$\cc\rightarrow\infty$ limit merely serves to reinstate the $\de$-functional 
charge constraint in the definition of the functional integral).  The 
interpretation of the new term is simple: it is simply a (divergent) spatial 
constant and is completely nonperturbative in origin.  As mentioned before, 
to make sense of the divergence, we shall consider $\cc$ as being finite 
until the last step of any calculation.

\section{\label{sec:dses}Dyson-Schwinger equations and truncation}
\setcounter{equation}{0}
Let us discuss the \DS equations and their truncation.  The techniques 
involved in the derivation of the \DS equations are standard, such that we 
shall present here only the most salient points in the interests of 
readability.  The reader is referred to 
Refs.~\cite{Watson:2006yq,Watson:2007vc,Popovici:2008ty} for an explicit 
account of the derivation of such \DS equations in Coulomb gauge.  Noticing 
that the Coulomb kernel occurring in the action cannot be written as a 
finite order polynomial in the fields (due to the presence of the 
\emph{inverse} Faddeev-Popov operator, this term is nonlocal), we must make 
some form of approximation in order to apply the standard \DS formalism.  To 
this end, we will introduce and justify a leading order truncation, whereby 
the Coulomb kernel occurring in the action is replaced by its expectation 
value and which will serve as an input into the resulting equations.

To start, let us consider the generating functional, \eq{eq:genfunc1}, in 
the absence of sources.  By implementing the transversality constraints on 
$\vec{A}$, $\vec{\pi}$ via Lagrange multiplier fields 
($\la,\ta$, respectively), the functional integral and corresponding action 
can be written
\be
Z=\int\cd\Phi e^{\imath\cs},\;\;\cs=\cs^2+\cs^3+\cs^4
\ee
where
\bea
\cs^2&=&\int dx\left\{
\ov{q}_{\al x}\left[\imath\ga^0\pd_{0x}
+\imath\s{\vec{\ga}}{\div_x}-m\right]_{\al\ba}q_{\ba x}
+\frac{1}{2}A_{ix}^a\left[\div_x^2\de_{ij}
-\nabla_{ix}\nabla_{jx}\right]A_{jx}^a
\right.\nonumber\\&&\left.
-\frac{1}{2}\pi_{ix}^a\pi_{ix}^a
+\pi_{ix}^a\pd_{0x}A_{ix}^a
-\la_x^a\nabla_{jx}A_{jx}^a
-\ta_x^a\nabla_{jx}\pi_{jx}^a\right\},
\nonumber\\
\cs^3&=&\int dx\,dy\,dz\,\left\{
-g\left[T^a\ga_i\right]_{\al\ba}\de(x-y)\de(x-z)
\ov{q}_{\al x}A_{iy}^aq_{\ba z}
+gf^{abc}\left[\nabla_{jx}\de(x-z)\right]\de(x-y)A_{kz}^aA_{jx}^bA_{ky}^c
\right\},
\nonumber\\
\cs^4&=&-\frac{1}{2}g^2\int dx\,dy
\left[f^{ade}A_{ix}^d\pi_{ix}^e
+\left[\ga^0T^a\right]_{\al\ba}\ov{q}_{\al x}q_{\ba x}\right]
\tilde{F}^{ab}\left(x,y;\vec{A}\right)
\left[f^{bfg}A_{jy}^f\pi_{jy}^g
+\left[\ga^0T^b\right]_{\ga\de}\ov{q}_{\ga y}q_{\de y}\right]
\nonumber\\
&&-\frac{1}{4}g^2f^{abc}f^{ade}\int dx\,A_{ix}^bA_{jx}^cA_{ix}^dA_{jx}^e,
\label{eq:actcomp0}
\eea
and where (recognizing the $\vec{A}$-dependence of the covariant derivative, 
$\vec{D}$, given by \eq{eq:covadj0})
\bea
\tilde{F}^{ab}\left(x,y;\vec{A}\right)&=&
\left[F^{ab}\left(x_0;\vec{x},\vec{y}\right)
+\cc\de^{ab}\right]\de(x_0-y_0)\nonumber\\
&=&
\left[-\s{\div_x}{\vec{D}_x^{ac}}\right]^{-1}\left(-\div_x^2\right)
\left[-\s{\div_x}{\vec{D}_x^{cb}}\right]^{-1}\de\left(x-y\right)
+\cc\de^{ab}\de(x_0-y_0).
\eea

Knowing that the \DS equations are formed via functional derivatives of the 
generating functional, we observe that the $\vec{A}$-dependence occurring 
within $\tilde{F}$ always comes with an associated factor of the coupling, 
such that explicit functional derivatives of $\tilde{F}$ will always result 
in additional loop structure.  As a leading (loop) order Ansatz, we 
therefore make the following truncation for the Coulomb kernel:
\be
\tilde{F}^{ab}\left(x,y;\vec{A}\right)\rightarrow
\tilde{F}^{ab}\left(x,y\right)
=\left[F(\vec{x}-\vec{y})+\cc\right]\de^{ab}\de(x_0-y_0)
\label{eq:akern}
\ee
where $F$ is some scalar function which will serve as a nonperturbative 
input to the system of \DS equations (the explicit expression will be 
discussed later).  The Coulomb kernel is instantaneous (as is its 
expectation value) so that in momentum space, the function $F$ will be 
independent of energy:
\be
\tilde{F}^{ab}\left(x,y\right)
=\int\dk{k}e^{-\imath k\cdot(x-y)}\tilde{F}^{ab}(k)
\sim\int\dk{k}e^{-\imath k\cdot(x-y)}\de^{ab}
\left[F(\vec{k}^2)+\cc(2\pi)^3\de(\vec{k})\right]
\label{eq:fdef0}
\ee
where $\dk{k}=d^4k/(2\pi)^4$.  It is clear from the discussion of the last 
section that $\tilde{F}$ is intimately related to the instantaneous part of 
the dressed temporal gluon propagator.  The tree-level contribution to 
$\tilde{F}$ is $\sim(-\div^2)^{-1}$ (or $1/\vec{k}^2$ in momentum space), 
i.e., independent of $\vec{A}$ and so the one-loop perturbative results are 
in principle preserved within this Ansatz.

Before continuing, it is worth briefly contrasting the formalism above with 
those of previous studies, namely 
Refs.~\cite{Watson:2006yq,Popovici:2008ty}.  These studies focused on a 
local form of the Coulomb gauge first order formalism, whereby the $\si$ 
and $\phi$ fields were not integrated out and the Faddeev-Popov determinant 
was written in terms of ghost fields.  This form is ideal for studying 
perturbation theory.  Here, after integrating out the $\si$ and $\phi$ 
fields, the Faddeev-Popov determinant cancels, leaving a form for the action 
involving the nonlocal Coulomb interaction term $\ro\tilde{F}\ro$.  The 
purely spatial gluonic and quark components of the action (i.e., those that 
exclusively involve only the $\vec{A}$, $\vec{\pi}$, $\ov{q}$ and $q$ 
fields) are unaltered, meaning that many of the previous results from 
Refs.~\cite{Watson:2006yq,Popovici:2008ty} pertaining to these components 
are retained.  After replacing $\tilde{F}$ with its expectation value 
(which serves as an external input into the system), the Coulomb interaction 
term involves three new momentum dependent tree-level interactions between 
the $\vec{A}$, $\vec{\pi}$ and quark fields (their explicit forms will be 
presented shortly).  These new interaction terms replace the dynamical 
interaction content of the $\si$, $\phi$ and ghost degrees of freedom with 
a simple set of effective vertices.  In effect, the full nonperturbative 
towers of \DS equations involving the $\si$, $\phi$ and ghost fields have 
been `solved' by the Ansatz for $\tilde{F}$.  Obviously, this `solution' is 
only a leading order Ansatz; we shall however see that important 
nonperturbative physics is nonetheless contained.  As has been seen, the 
removal of the $\si$, $\phi$ and ghost fields in the nonlocal formalism is 
intimately related to the imposal of Gauss' law and (total) charge 
conservation; in the local formulation, the Slavnov-Taylor identities 
perform this role \cite{Watson:2008fb}.

The \DS equations are integral equations relating the various proper 
(one-particle irreducible, [1PI]) Green's functions of the theory.  The 
most basic of the 1PI functions are the two-point proper functions and in 
the current formalism, with only $\vec{A}$, $\vec{\pi}$ and quark fields 
(aside from the trivial Lagrange multiplier fields), the most general 
momentum space decomposition of these is given by (see 
Refs.~\cite{Watson:2006yq,Popovici:2008ty} for details of the derivation 
and notation)
\bea
\G_{\pi\pi ij}^{ab}(k)&=&\imath\de^{ab}\left[\de_{ij}\G_{\pi\pi}(k)
+l_{ij}(\vec{k})\ov{\G}_{\pi\pi}(k)\right],\nonumber\\
\G_{A\pi ij}^{ab}(k)&=&\de^{ab}k_0\left[\de_{ij}\G_{A\pi}(k)
+l_{ij}(\vec{k})\ov{\G}_{A\pi}(k)\right]=\G_{\pi Aij}^{ab}(-k),\nonumber\\
\G_{AAij}^{ab}(k)&=&\imath\de^{ab}\vec{k}^2\left[t_{ij}(\vec{k})\G_{AA}(k)
+l_{ij}(\vec{k})\ov{\G}_{AA}(k)\right],\nonumber\\
\G_{\ov{q}q\al\ba}^{(0)}(k)&=&\imath\left[\ga^0k_0A_t(k)
-\s{\vec{\ga}}{\vec{k}}A_s(k)-B_m(k)
+\ga^0k_0\s{\vec{\ga}}{\vec{k}}A_d(k)\right]_{\al\ba}
\label{eq:gdecomp0}
\eea
where $l_{ij}(\vec{k})=k_ik_j/\vec{k}^2$ is the longitudinal spatial 
projector and $t_{ij}(\vec{k})=\de_{ij}-l_{ij}(\vec{k})$ is the transverse 
spatial projector.  All (scalar and dimensionless with the exception of 
$B_m$ and $A_d$) dressing functions are functions of $k_0^2$ and $\vec{k}^2$ 
separately due to the inherent noncovariance of Coulomb gauge.  At 
tree-level, the dressing functions reduce to (the tree-level forms for the 
proper two-point functions follow directly from the quadratic part of the 
action, $\cs^2$, given in \eq{eq:actcomp0})
\be
\G_{AA}=\G_{A\pi}=\G_{\pi\pi}=A_t=A_s=1,\;\;\;\;B_m=m,\;\;\;\;\ov{\G}_{AA}
=\ov{\G}_{A\pi}=\ov{\G}_{\pi\pi}=A_d=0.
\ee

Alongside the proper two-point functions, one is also interested in the 
corresponding propagators (connected two-point Green's functions).  The 
connection between the connected and proper Green's functions is supplied 
via the Legendre transform.  Since the components of the gluon field are 
treated individually in Coulomb gauge, the gluon propagator dressing 
function is not simply the inverse of the corresponding proper dressing 
function (as in Landau gauge) but rather, a matrix inversion structure 
arises.  Taking into account the Lagrange multiplier fields that enforce 
the transversality of the $\vec{A}$ and $\vec{\pi}$-fields, the components 
of the gluon propagator are given by (see also Ref.~\cite{Watson:2006yq})
\bea
W_{AAij}^{ab}(k)&=&
\imath\de^{ab}t_{ij}(\vec{k})\frac{\G_{\pi\pi}(k)}{\Delta_g(k)},\nonumber\\
W_{A\pi ij}^{ab}(k)&=&
-\de^{ab}k_0t_{ij}(\vec{k})\frac{\G_{A\pi}(k)}{\Delta_g(k)},\nonumber\\
W_{\pi\pi ij}^{ab}(k)&=&
\imath\de^{ab}\vec{k}^2t_{ij}(\vec{k})\frac{\G_{AA}(k)}{\Delta_g(k)}
\label{eq:wgdecomp0}
\eea
where the common denominator factor, $\Delta_g(k)$, including the Feynman 
prescription, is given by
\be
\Delta_g(k)=k_0^2\G_{A\pi}^2(k)-\vec{k}^2\G_{AA}(k)\G_{\pi\pi}(k)+\imath0_+.
\label{eq:deltag0}
\ee
In the case of the quarks (see Ref.~\cite{Popovici:2008ty}) we have
\be
W_{\ov{q}q\al\ba}(k)=-\frac{\imath}{\Delta_f(k)}\left[\ga^0k_0A_t(k)
-\s{\vec{\ga}}{\vec{k}}A_s(k)+B_m(k)
+\ga^0k_0\s{\vec{\ga}}{\vec{k}}A_d(k)\right]_{\al\ba}
\label{eq:wqdecomp0}
\ee
where the denominator factor is given by
\be
\Delta_f(k)=
k_0^2A_t^2(k)-\vec{k}^2A_s^2(k)-B_m^2(k)+k_0^2\vec{k}^2A_d^2(k)+\imath0_+.
\label{eq:deltaf0}
\ee
The matrix inversion structure for the gluon and quark propagators will have 
important consequences when solving the \DS equations.  At tree-level, the 
denominator structures reduce to the familiar forms: 
$\Delta_g=k_0^2-\vec{k}^2+\imath0_+$ and 
$\Delta_f=k_0^2-\vec{k}^2-m^2+\imath0_+$.

The tree-level vertex (three- and four-point proper) functions can be 
derived from the cubic and quartic parts of the action ($\cs^3$ and 
$\cs^4$ of \eq{eq:actcomp0}, respectively).  As for the two-point functions, 
the purely spatial tree-level vertex functions of the local first order 
formalism can be taken from Refs.~\cite{Watson:2006yq,Popovici:2008ty}.  For 
these, the momentum space expressions are
\bea
\G_{AAAijk}^{(0)abc}(k_1,k_2,k_3)&=&
-\imath gf^{abc}\left[\de_{ij}(k_1-k_2)_k+\de_{jk}(k_2-k_3)_i
+\de_{ki}(k_3-k_1)_j\right],\nonumber\\
\G_{AAAAijkl}^{(0)abcd}(k_1,k_2,k_3,k_4)&=&
-\imath g^2\left[\de_{ij}\de_{kl}(f^{ace}f^{bde}-f^{ade}f^{cbe})
+\de_{ik}\de_{jl}(f^{abe}f^{cde}-f^{ade}f^{bce})
\right.\nonumber\\&&\left.
+\de_{il}\de_{jk}(f^{ace}f^{dbe}-f^{abe}f^{cde})\right],\nonumber\\
\G_{\ov{q}qA\al\ba i}^{(0)a}(k_1,k_2,k_3)&=&
-g\left[\ga^iT^a\right]_{\al\ba},
\label{eq:treev0}
\eea
where it is understood that all momenta are incoming and energy-momentum 
conservation has been applied ($\sum k_i=0$).  The new interaction terms 
that arise in the present nonlocal formalism, with the Ansatz that the 
Coulomb kernel be replaced by its expectation value, are all contained 
within the quartic component of the action ($\cs^4$ of \eq{eq:actcomp0}).  
They are all linear in $\tilde{F}$ and since the color charge ($\hat{\ro}$) 
involves both gluonic and quark components on the same footing, the 
tree-level vertices all have the same structure.  It will be seen later that 
indeed, the resulting \DS equations for the gluon and quark sectors have 
very similar forms.  Using the techniques of 
Refs.~\cite{Watson:2006yq,Popovici:2008ty}, the momentum space expressions 
for the vertices are:
\bea
\G_{AA\pi\pi ijkl}^{(0)abcd}(k_1,k_2,k_3,k_4)&=&
-\imath g^2\left[f^{ead}f^{fbc}\de_{il}\de_{jk}\tilde{F}^{ef}(k_1+k_4)
+f^{ebd}f^{fac}\de_{jl}\de_{ik}\tilde{F}^{ef}(k_1+k_3)\right],\nonumber\\
\G_{\ov{q}qA\pi\al\ba ij}^{(0)ab}(k_1,k_2,k_3,k_4)&=&
\imath g^2f^{abe}\left[\ga^0T^f\right]_{\al\ba}\de_{ij}
\tilde{F}^{ef}(k_1+k_2),\nonumber\\
\G_{\ov{q}q\ov{q}q\al\ba\ga\de}^{(0)}(k_1,k_2,k_3,k_4)&=&
-\imath g^2\left[\ga^0T^a\right]_{\al\ba}\left[\ga^0T^b\right]_{\ga\de}
\tilde{F}^{ab}(k_1+k_2)
+\imath g^2\left[\ga^0T^a\right]_{\al\de}\left[\ga^0T^b\right]_{\ga\ba}
\tilde{F}^{ba}(k_1+k_4).
\label{eq:treev1}
\eea
Notice that the above vertices are written such that the symmetry 
properties are manifest.

Let us now discuss the \DS equations themselves.  As stated earlier, we 
shall not present details of their derivation here: the basic techniques of 
such a derivation in Coulomb gauge are expounded in 
Refs.~\cite{Watson:2006yq,Popovici:2008ty}.  Generically, the \DS equations 
have a very definite structure of loop terms that arises from the 
combination of repeated functional differentiation of the generating 
functional and the Legendre transform connecting the connected and proper 
Green's functions.  Once the notation and conventions have been established, 
different interaction terms merely follow this characteristic pattern; the 
difficulty is simply in keeping track of the various coefficients, 
especially where anticommuting Grassmann-valued fields (quarks or ghosts) 
are present.

\begin{figure}[t]
\vspace{0.8cm}
\includegraphics[width=0.7\linewidth]{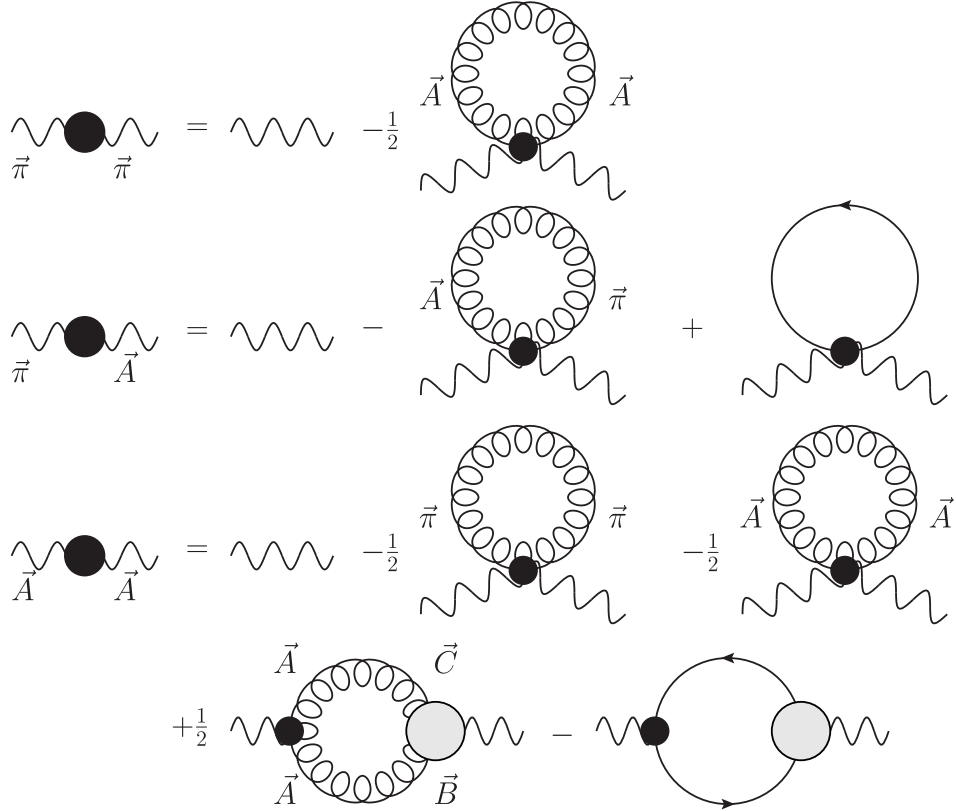}
\vspace{0.3cm}
\caption{\label{fig:gdse0}\DS equations for $\G_{\pi\pi}$, $\G_{\pi A}$ 
and $\G_{AA}$, omitting two-loop terms.  Wavy lines denote proper functions, 
the large filled blob indicating the dressed function.  Springs denote 
gluonic propagators, lines denote the quark propagator and all internal 
propagators are dressed.  Small blobs indicate tree-level vertices and large 
circles denote dressed vertices.  The gluonic field types $\vec{B}$ and 
$\vec{C}$ denote the sum over $\vec{A}$ and $\vec{\pi}$ contributions 
arising due to the presence of mixed gluon propagators.  See text for 
details.}
\end{figure}

\begin{figure}[t]
\vspace{0.8cm}
\includegraphics[width=0.6\linewidth]{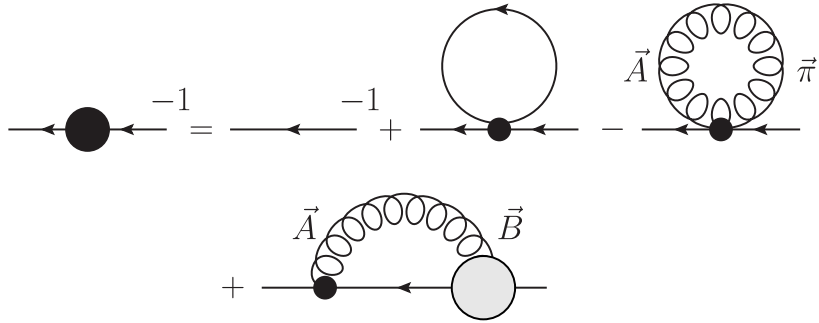}
\vspace{0.3cm}
\caption{\label{fig:qdse0}\DS equation for the quark two-point function, 
omitting two-loop terms.  On the left-hand side, the filled blob indicates 
the dressed (inverse) propagator.  Lines denote quark propagators, springs 
denote gluon propagators and all internal propagators are dressed.  Small 
blobs indicate tree-level vertices and large circles denote dressed 
vertices.  The gluonic field type $\vec{B}$ represents the sum over 
$\vec{A}$ and $\vec{\pi}$ contributions arising due to the presence of 
mixed gluon propagators.  See text for details.}
\end{figure}

The \DS equations for the two-point proper functions (in momentum space) 
are presented below.  To aid presentation, the two-loop terms that will not 
explicitly be used in this study are omitted (they are collectively denoted 
$\G^{(2)}$ below).  The equations are
\bea
\G_{\pi\pi ij}^{ab}(k)&=&\G_{\pi\pi ij}^{(0)ab}(k)
-\frac{1}{2}
\int\dk{\w}\G_{AA\pi\pi klij}^{(0)cdab}(\w,-\w,k,-k)W_{AAlk}^{dc}(\w)
+\G_{\pi\pi ij}^{(2)ab}(k),
\label{eq:dsepp0}
\\
\G_{\pi Aij}^{ab}(k)&=&\G_{\pi Aij}^{(0)ab}(k)
-\int\dk{\w}\G_{AA\pi\pi kjli}^{(0)cbda}(-\w,-k,\w,k)W_{A\pi kl}^{cd}(\w)
\nonumber\\&&
+\int\dk{\w}\G_{\ov{q}qA\pi\al\ba ji}^{(0)ba}(\w,-\w,-k,k)
W_{\ov{q}q\ba\al}(\w)
+\G_{\pi Aij}^{(2)ab}(k),
\label{eq:dsepa0}
\\
\G_{AAij}^{ab}(k)&=&\G_{AAij}^{(0)ab}(k)
-\frac{1}{2}\int\!\dk{\w}\G_{AA\pi\pi ijkl}^{(0)abcd}(k,-k,\w,-\w)
W_{\pi\pi lk}^{dc}(\w)
-\frac{1}{2}\int\!\dk{\w}\G_{AAAAijkl}^{(0)abcd}(k,-k,\w,-\w)
W_{AA lk}^{dc}(\w)
\nonumber\\&&
+\frac{1}{2}\int\dk{\w}\G_{AAAikl}^{(0)acd}(k,-\w,\w-k)W_{ABkk'}^{cc'}(\w)
W_{ACll'}^{dd'}(k-\w)\G_{CBAl'k'j}^{d'c'b}(k-\w,\w,-k)
\nonumber\\&&
-\int\dk{\w}\G_{\ov{q}qA\al\ba i}^{(0)a}(\w-k,-\w,k)W_{\ov{q}q\ba\ba'}(\w)
\G_{\ov{q}qA\ba'\al' j}^{b}(\w,k-\w,-k)W_{\ov{q}q\al'\al}(\w-k)
+\G_{AAij}^{(2)ab}(k),
\label{eq:dseaa0}
\eea
\bea
\G_{\ov{q}q\al\ba}(k)&=&\G_{\ov{q}q\al\ba}^{(0)}(k)
+\int\dk{\w}\G_{\ov{q}q\ov{q}q\al\ba\ga\de}^{(0)}(k,-k,\w,-\w)
W_{\ov{q}q\de\ga}(\w)
-\int\dk{\w}\G_{\ov{q}qA\pi\al\ba kl}^{(0)cd}(k,-k,\w,-\w)W_{\pi Alk}^{dc}(\w)
\nonumber\\&&
+\int\dk{\w}\G_{\ov{q}qA\al\al' k}^{(0)c}(k,-\w,\w-k)W_{\ov{q}q\al'\ba'}(\w)
W_{ABkk'}^{cc'}(k-\w)\G_{\ov{q}qB\ba'\ba k'}^{c'}(\w,-k,k-\w)
\nonumber\\&&
+\G_{\ov{q}q\al\ba}^{(2)}(k).
\label{eq:dseqq0}
\eea
Because of the existence of the mixed gluon propagator $W_{A\pi}$ and the 
possibility of dressed vertices such as $\G_{\pi\pi A}$ or 
$\G_{\ov{q}q\pi}$, in certain terms of the above equations (in particular, 
the one-loop term of \eq{eq:dseaa0} involving the three-gluon vertex and 
the spatial one-loop component of the quark self-energy in the above) one 
must sum up over the possible gluonic field types $\vec{A}$ and $\vec{\pi}$: 
this sum is denoted by the repeated subscript indices 
$\vec{B},\vec{C},\ldots$ (obviously, no relation to the chromomagnetic or 
ghost fields).  The equations are presented diagrammatically in 
Figs.~\ref{fig:gdse0} and \ref{fig:qdse0}.  As mentioned previously, the 
Ansatz of replacing the Coulomb kernel with its expectation value in the 
action does not interfere with the one-loop \DS equations.  Inserting the 
appropriate tree-level vertices and propagators into the above equations 
and using
\be
\tilde{F}^{ab}(k)=\de^{ab}\frac{1}{\vec{k}^2}
\ee
(equivalent to the tree-level temporal gluon propagator), it can indeed be 
verified that the known one-loop perturbative expressions 
\cite{Watson:2007mz,Popovici:2008ty} are recovered.

To complete this section, let us now describe the remainder of the 
truncation scheme.  We are interested here in the leading loop, 
nonperturbative infrared behavior; in particular, on the effect of the 
terms generated directly by the Coulomb kernel, $\tilde{F}$.  As discussed 
earlier, when replacing the full (nonlocal) Coulomb kernel with its 
expectation value within the action, one has already truncated out certain 
higher loop terms.  The resulting \DS equations retain their full leading 
(one-loop perturbative) structure.  The next truncation we make on the \DS 
equations is thus to restrict to the one-loop terms.

The next part of the truncation scheme is to omit those terms generated by 
the tree-level $\G_{AAA}^{(0)}$, $\G_{AAAA}^{(0)}$ and $\G_{\ov{q}qA}^{(0)}$ 
vertices.  These are the terms that are present at one-loop in the \DS 
equations, but do not involve the Coulomb kernel directly.  What remains of 
the \DS equations are the tadpole terms involving the \emph{tree-level} 
four-point vertices $\G_{AA\pi\pi}^{(0)}$, $\G_{\ov{q}qA\pi}^{(0)}$ and 
$\G_{\ov{q}q\ov{q}q}^{(0)}$: these terms explicitly involve only the Coulomb 
kernel $\tilde{F}$ and the propagators, thus forming a closed set of 
equations for a given input function $\tilde{F}$.

In order to complete the truncation scheme for the \DS equations, we must 
supply an expression for the Coulomb kernel.  This is based on an infrared 
divergent $1/\vec{k}^4$ behavior.  In this study, we shall not include the 
perturbative ($1/\vec{k}^2$) term, since we are primarily interested in the 
infrared; however, an interesting point of note is that logarithmic 
ultraviolet divergences will still emerge (this will be discussed later).  
The $1/\vec{k}^4$ factor is justified on two grounds: lattice studies do 
show this behavior for the instantaneous component of the temporal gluon 
propagator, although the extant results admittedly do not go far enough 
into the infrared regime for this to be definitive \cite{Nakagawa:2011ar,
Iritani:2011zg,Nakagawa:2009is,Quandt:2008zj,Cucchieri:2007uj,
Langfeld:2004qs,Cucchieri:2000gu}.  The second reason is somewhat more 
pragmatic: as will be seen, such an infrared behavior is not only consistent 
with a linearly rising potential between heavy quarks but also generates 
dynamical chiral symmetry breaking for chiral quarks.  Further, the product 
$g^2\tilde{F}$ is a renormalization group invariant quantity 
\cite{Cucchieri:2000hv,Zwanziger:1998ez} and serves as an ideal quantity to 
use as input.  For technical reasons, we use two forms for the input Coulomb 
kernel.  The expressions are:
\bea
g^2C_F\tilde{F}^{ab}(k)&=&\de^{ab}{\cal C}(2\pi)^3\de(\vec{k})
+\de^{ab}F(\vec{k}^2),\nonumber\\
F(\vec{k}^2)&=&8\pi\si_c\left\{\begin{array}{cc}\frac{1}{[\vec{k}^2]^2}&
\mbox{(gluon/quark DSe)},\\\frac{1}{[\vec{k}^2+\xi]^2}&\mbox{(quark DSe)},
\end{array}\right.
\label{eq:tf0}
\eea
where the coefficients have been chosen for later convenience 
($C_F=(N_c^2-1)/2N_c$ is a color factor).  The coefficient $\si_c$ is the 
Coulomb string tension (as distinct from the physical Wilson string tension, 
see e.g., Refs.~\cite{Iritani:2011zg,Zwanziger:2002sh} for a discussion).  
In the \DS equations, an unadulterated $1/\vec{k}^4$ factor results in an 
infrared divergence, so that we will introduce a numerical infrared cutoff 
regulator.  In the case of the quark, a second infrared mass regulator, 
$\xi$, will be considered --- this form of regularization will turn out to 
have a very clear physical interpretation, which we will discuss later.  
Recall that the constant ${\cal C}$ multiplying the $\de$-function (arising 
from the total charge constraint) is considered to be finite until the end 
of the calculation whereupon we consider the limit 
${\cal C}\rightarrow\infty$.

\section{\label{sec:andev}Truncated \DS equations: analytic development}
\setcounter{equation}{0}
Let us now take the truncated \DS equations and decompose them into a form 
useful for further analysis.

\subsection{$\G_{\pi A}$ equation}
The truncated \DS equation for $\G_{\pi A}$, \eq{eq:dsepa0}, reads
\bea
\G_{\pi Aij}^{ab}(k)&=&\G_{\pi Aij}^{(0)ab}(k)
-\int\dk{\w}\G_{AA\pi\pi kjli}^{(0)cbda}(-\w,-k,\w,k)W_{A\pi kl}^{cd}(\w)
\nonumber\\&&
+\int\dk{\w}\G_{\ov{q}qA\pi\al\ba ji}^{(0)ba}(\w,-\w,-k,k)
W_{\ov{q}q\ba\al}(\w).
\eea
Inserting the tree-level vertex functions using \eq{eq:treev1} and the 
general forms for the propagators and proper two-point function, 
Eqs.~(\ref{eq:wgdecomp0},\ref{eq:gdecomp0}), one obtains
\bea
\lefteqn{
-\de^{ab}k_0\left[\de_{ij}\G_{A\pi}(k)
+l_{ij}(\vec{k})\ov{\G}_{A\pi}(k)\right]=-\de^{ab}k_0\de_{ij}
}\nonumber\\&&
-\imath g^2\int\dk{\w}\left[f^{eca}f^{ebd}\de_{ki}\de_{jl}\tilde{F}(k-\w)
+f^{eba}f^{ecd}\de_{ji}\de_{kl}\tilde{F}(\w-\w)\right]
\de^{cd}\w_0t_{kl}(\vec{\w})\frac{\G_{A\pi}(\w)}{\Delta_g(\w)}
\nonumber\\&&
+g^2f^{aeb}\left[\ga^0T^e\right]_{\al\ba}\de_{ji}
\int\frac{\dk{\w}}{\Delta_f(\w)}\tilde{F}(\w-\w)
\left[\ga^0\w_0A_t(\w)-\s{\vec{\ga}}{\vec{\w}}A_s(\w)+B_m(\w)
+\ga^0\w_0\s{\vec{\ga}}{\vec{\w}}A_d(\w)\right]_{\ba\al}.
\eea
We notice the occurrence of factors $\tilde{F}(\w-\w)$, which are 
technically undefined.  However, the corresponding color structures 
read $f^{ecd}\de^{cd}$ and $[T^e]_{\al\ba}\de_{\ba\al}$, both of which 
are automatically zero.  Assuming that $\tilde{F}$ is infrared regulated, 
these terms would then pose no problem and we summarily dismiss them.  
Expanding the denominator factor $\Delta_g$ using \eq{eq:deltag0}, the 
equation then reads
\be
\de_{ij}\G_{A\pi}(k)+l_{ij}(\vec{k})\ov{\G}_{A\pi}(k)=\de_{ij}
-\imath g^2N_c\int
\frac{\dk{\w}\w_0\G_{A\pi}(\w)}{k_0\left[\w_0^2\G_{A\pi}^2(\w)
-\vec{\w}^2\G_{AA}(\w)\G_{\pi\pi}(\w)+\imath0_+\right]}t_{ij}(\vec{\w})
\tilde{F}(k-\w).
\ee
Since $\tilde{F}$ is independent of the energy and the scalar dressing 
functions $\G_{AA}$, $\G_{A\pi}$, $\G_{\pi\pi}$ are all even functions of 
energy, the integrand is overall odd in the energy and thus vanishes without 
further assumption.  We thus have the nonperturbative result that under the 
current truncation scheme,
\be
\G_{A\pi}(k)=1,\;\;\ov{\G}_{A\pi}(k)=0.
\ee
In other words, the mixed gluon two-point proper function remains bare.  
This result is very useful because as will be seen below, the energy 
dependence of the gluon propagators turns out to be trivial.

\subsection{$\G_{\pi\pi}$ and $\G_{AA}$: the gluon gap equation}
The truncated \DS equations for $\G_{\pi\pi}$ and $\G_{AA}$, 
Eqs.~(\ref{eq:dsepp0},\ref{eq:dseaa0}), can be treated simultaneously.  
They read
\bea
\G_{\pi\pi ij}^{ab}(k)&=&
\G_{\pi\pi ij}^{(0)ab}(k)
-\frac{1}{2}\int\dk{\w}\G_{AA\pi\pi klij}^{(0)cdab}(\w,-\w,k,-k)
W_{AAlk}^{dc}(\w),
\nonumber\\
\G_{AAij}^{ab}(k)&=&
\G_{AAij}^{(0)ab}(k)
-\frac{1}{2}\int\dk{\w}\G_{AA\pi\pi ijkl}^{(0)abcd}(k,-k,\w,-\w)
W_{\pi\pi lk}^{dc}(\w).
\eea
Notice that having restricted the truncation scheme to include only those 
one-loop terms involving the Coulomb kernel, the two equations are identical 
in their structure.  Inserting the tree-level vertex functions, 
\eq{eq:treev1}, the two-point function decompositions, 
Eqs.~(\ref{eq:gdecomp0},\ref{eq:wgdecomp0},\ref{eq:deltag0}), projecting 
onto the transverse components (the longitudinal components of the proper 
two-point functions will play no role) and using the previous result that 
$\G_{A\pi}=1$ within this truncation scheme, one readily obtains
\bea
\G_{\pi\pi}(k)&=&1
+\frac{\imath}{2}g^2N_c\int\frac{\dk{\w}\G_{\pi\pi}(\w)}{\left[\w_0^2
-\vec{\w}^2\G_{AA}(\w)\G_{\pi\pi}(\w)+\imath0_+\right]}
\tilde{F}(k-\w)t_{ji}(\vec{k})t_{ij}(\vec{\w}),\nonumber\\
\G_{AA}(k)&=&1
+\frac{\imath}{2}g^2N_c\int
\frac{\dk{\w}\vec{\w}^2\G_{AA}(\w)}{\vec{k}^2\left[\w_0^2
-\vec{\w}^2\G_{AA}(\w)\G_{\pi\pi}(\w)+\imath0_+\right]}
\tilde{F}(k-\w)t_{ji}(\vec{k})t_{ij}(\vec{\w}).
\eea
Given that $\tilde{F}$ is energy independent, the energy integrals of the 
above are relatively trivial and since there is no $k_0$-dependence, the 
proper dressing functions are energy independent.  However, we should point 
out that really, the energy integral is only trivial if the spatial 
functions are regularized and finite.  Here, this refers to an implicit 
infrared regularization in the case of $\tilde{F}\sim1/\vec{k}^4$ with its 
strong infrared singularity, the finite coefficient ${\cal C}$ multiplying 
$\de(\vec{k})$ and more generally, the ultraviolet cutoff when one considers 
the perturbative term $\tilde{F}\sim1/\vec{k}^2$ although we shall see that 
the $\G_{AA}$ equation involves a logarithmic UV divergence even with the 
$1/\vec{k}^4$ interaction.  It is helpful to define the static propagators 
$W^{(s)}$ as the energy integral of the full propagators (in configuration 
space, these are the equaltime propagators).  In the case of the $W_{AA}$ 
propagator,
\be
W_{AAij}^{(s)ab}(\vec{k})
=\imath\de^{ab}t_{ij}(\vec{k})\int_{-\infty}^{\infty}\frac{dk_0}{2\pi}
\frac{\G_{\pi\pi}(\vec{k}^2)}{\left[k_0^2
-\vec{k}^2\G_{AA}(\vec{k}^2)\G_{\pi\pi}(\vec{k}^2)+\imath0_+\right]}
=\de^{ab}t_{ij}(\vec{k})\int_{-\infty}^{\infty}\frac{dk_4}{2\pi}
\frac{\G_{\pi\pi}(\vec{k}^2)}{\left[k_4^2+\vec{k}^2\G_{AA}(\vec{k}^2)
\G_{\pi\pi}(\vec{k}^2)\right]}
\ee
where in the second integral form, a Wick rotation 
($k_0\rightarrow\imath k_4$) has been performed.  Doing the integral, one 
finds that the static propagator $W_{AA}^{(s)}$ can be written in terms of 
a single dressing function, which we denote $G$:
\be
W_{AAij}^{(s)ab}(\vec{k})
=\de^{ab}t_{ij}(\vec{k})\frac{1}{2|\vec{k}|}G(\vec{k}^2)^{1/2},\;\;\;\;
G(\vec{k}^2)=\frac{\G_{\pi\pi}(\vec{k}^2)}{\G_{AA}(\vec{k}^2)}.
\label{eq:grib0}
\ee
The static propagator $W_{\pi\pi}^{(s)}$ can also be written in terms of $G$:
\be
W_{\pi\pi ij}^{(s)ab}(\vec{k})
=\de^{ab}t_{ij}(\vec{k})\frac{|\vec{k}|}{2}G(\vec{k}^2)^{-1/2}.
\ee
The reduction of the two dressing functions $\G_{AA}$ and $\G_{\pi\pi}$ to 
a single function $G$ follows directly from the previous result $\G_{A\pi}=1$.

Returning to the \DS equations, we can now write
\bea
\G_{\pi\pi}(\vec{k}^2)&=&1
+\frac{1}{4}g^2N_c\int\frac{\dk{\vec{\w}}}{\sqrt{\vec{\w}^2}}
G(\vec{\w}^2)^{1/2}\tilde{F}(k-\w)t_{ji}(\vec{k})t_{ij}(\vec{\w}),\nonumber\\
\G_{AA}(\vec{k}^2)&=&1
+\frac{1}{4}g^2N_c\int\frac{\dk{\vec{\w}}}{\sqrt{\vec{\w}^2}}
\frac{\vec{\w}^2}{\vec{k}^2}G(\vec{\w}^2)^{-1/2}
\tilde{F}(k-\w)t_{ji}(\vec{k})t_{ij}(\vec{\w})
\eea
where $\dk{\vec{\w}}=d\vec{\w}/(2\pi)^3$.  Expanding out $\tilde{F}$ with 
\eq{eq:tf0} we have
\bea
\G_{\pi\pi}(\vec{k}^2)&=&1
+\frac{1}{2}\frac{N_c}{C_F}\frac{{\cal C}}{\sqrt{\vec{k}^2}}G(\vec{k}^2)^{1/2}
+\frac{1}{4}\frac{N_c}{C_F}\int\frac{\dk{\vec{\w}}}{\sqrt{\vec{\w}^2}}
G(\vec{\w}^2)^{1/2}F(\vec{k}-\vec{\w})t_{ji}(\vec{k})t_{ij}(\vec{\w}),
\nonumber\\
\G_{AA}(\vec{k}^2)&=&1+\frac{1}{2}\frac{N_c}{C_F}
\frac{{\cal C}}{\sqrt{\vec{k}^2}}G(\vec{k}^2)^{-1/2}
+\frac{1}{4}\frac{N_c}{C_F}\int\frac{\dk{\vec{\w}}}{\sqrt{\vec{\w}^2}}
\frac{\vec{\w}^2}{\vec{k}^2}G(\vec{\w}^2)^{-1/2}F(\vec{k}-\vec{\w})
t_{ji}(\vec{k})t_{ij}(\vec{\w}).
\eea
The proper dressing functions, $\G_{\pi\pi}$ and $\G_{AA}$ (and 
subsequently, the corresponding propagators), are explicitly dependent not 
only on $\cc$, but also include potentially infrared divergent contributions 
stemming from the integrals involving $F$.  However, using the definition of 
$G$, one can easily see that the above equations can be written in terms of 
a single equation for $G$:
\be
G(\vec{k}^2)=1
+\frac{1}{4}\frac{N_c}{C_F}\int\frac{\dk{\vec{\w}}}{\sqrt{\vec{\w}^2}}
F(\vec{k}-\vec{\w})t_{ji}(\vec{k})t_{ij}(\vec{\w})\left[G(\vec{\w}^2)^{1/2}
-\frac{\vec{\w}^2}{\vec{k}^2}\frac{G(\vec{k}^2)}{G(\vec{\w}^2)^{1/2}}\right].
\label{eq:ggap0}
\ee
The terms proportional to ${\cal C}$ cancel, showing that $G$ and the static 
propagators are independent of ${\cal C}$.  In addition, one sees that the 
infrared divergence of $F$ is tempered, such that $G$ may be infrared finite 
(as will be seen).  It would thus appear that the physical dynamics are 
contained within the static propagator dressing function, whereas the full 
propagators (and in particular, their pole positions) are not physical.  We 
shall discuss this at the end.  For reasons that will become obvious 
shortly, we shall refer to \eq{eq:ggap0} as the gluon gap equation.  The 
form of the equation is identical to that for the static gluon propagator 
derived from the canonical approach \cite{Szczepaniak:2001rg} (see also the 
earlier work of Refs.~\cite{Schutte:1985sd,Szczepaniak:1995cw}).  In the 
canonical approach, the Coulomb kernel $F$ (here an input) is represented by 
a combination of ghost dressing and Coulomb form factors, which are 
self-consistently determined from their respective equations.  Later work 
within the canonical approach included further terms: the ghost `curvature' 
\cite{Szczepaniak:2003ve,Feuchter:2004mk,Reinhardt:2004mm}, and three- and 
four-gluon interactions \cite{Campagnari:2010wc}.  The equivalence of the 
truncated \DS equation above, \eq{eq:ggap0}, to its counterpart arising from 
the canonical formalism and considered in Ref.~\cite{Szczepaniak:2001rg} is 
one of the results of this paper.

\subsection{$\G_{\ov{q}q}$: the quark gap equation}
Under truncation, the \DS equation for $\G_{\ov{q}q}$, \eq{eq:dseqq0}, reads
\be
\G_{\ov{q}q\al\ba}(k)=\G_{\ov{q}q\al\ba}^{(0)}(k)
+\int\dk{\w}\G_{\ov{q}q\ov{q}q\al\ba\ga\de}^{(0)}(k,-k,\w,-\w)
W_{\ov{q}q\de\ga}(\w)
-\int\dk{\w}\G_{\ov{q}qA\pi\al\ba kl}^{(0)cd}(k,-k,\w,-\w)
W_{\pi Alk}^{dc}(\w).
\ee
Inserting the appropriate tree-level vertices, \eq{eq:treev1} and 
propagators, Eqs.~(\ref{eq:wgdecomp0},\ref{eq:wqdecomp0}), resolving the 
color algebra (again discarding terms where one has $\mbox{Tr}[T^a]$ or 
$f^{cac}$ that multiply $\tilde{F}(\w-\w)$ as previously discussed for 
$\G_{A\pi}$), one obtains
\be
\G_{\ov{q}q\al\ba}(k)=\G_{\ov{q}q\al\ba}^{(0)}(k)
+g^2C_F\int\frac{\dk{\w}\tilde{F}(k-\w)}{\Delta_f(\w)}
\left[\ga^0\w_0A_t(\w)-\ga^0\s{\vec{\ga}}{\vec{\w}}\ga^0A_s(\w)+B_m
+\s{\vec{\ga}}{\vec{\w}}\ga^0\w_0A_d(\w)\right]_{\al\ba},
\ee
where it is again recognized that $\tilde{F}$ is independent of the energy.  
Projecting out the Dirac components in the decomposition for $\G_{\ov{q}q}$, 
\eq{eq:gdecomp0}, one obtains four equations for the dressing functions:
\bea
A_t(k)&=&1
-\imath g^2C_F\int\frac{\dk{\w}\w_0A_t(\w)\tilde{F}(k-\w)}{k_0\Delta_f(\w)},\\
A_s(k)&=&1
+\imath g^2C_F\int\frac{\dk{\w}\s{\vec{k}}{\vec{\w}}A_s(\w)
\tilde{F}(k-\w)}{\vec{k}^2\Delta_f(\w)},\\
B_m(k)&=&m
+\imath g^2C_F\int\frac{\dk{\w}B_m(\w)\tilde{F}(k-\w)}{\Delta_f(\w)},\\
A_d(k)&=&
\imath g^2C_F\int\frac{\dk{\w}\w_0\s{\vec{k}}{\vec{\w}}A_d(\w)
\tilde{F}(k-\w)}{k_0\vec{k}^2\Delta_f(\w)}.
\eea
As for the $\G_{A\pi}$ equation, the odd energy integrals vanish since all 
dressing functions are functions of $\w_0^2$ (including $\Delta_f$), 
furnishing the result that
\be
A_t(k)=1,\;\;\;\;A_d(k)=0.
\ee
In addition, one sees that $A_s$ and $B_m$ are independent of energy.  The 
static quark propagator, defined in analogy to the gluon propagator, is then
\be
W_{\ov{q}q\al\ba}^{(s)}(\vec{k})=
-\imath\int_{-\infty}^{\infty}\frac{dk_0}{2\pi}
\frac{\left[\ga^0k_0-\s{\vec{\ga}}{\vec{k}}A_s(\vec{k}^2)
+B_m(\vec{k}^2)\right]_{\al\ba}}{\left[k_0^2-\vec{k}^2A_s^2(\vec{k}^2)
-B_m^2(\vec{k}^2)+\imath0_+\right]}
=\int_{-\infty}^{\infty}\frac{dk_4}{2\pi}
\frac{\left[\s{\vec{\ga}}{\vec{k}}A_s(\vec{k}^2)
-B_m(\vec{k}^2)\right]_{\al\ba}}{\left[k_4^2+\vec{k}^2A_s^2(\vec{k}^2)
+B_m^2(\vec{k}^2)\right]}
\ee
and performing the integral, one obtains
\be
W_{\ov{q}q\al\ba}^{(s)}(\vec{k})=\frac{\left[\s{\vec{\ga}}{\vec{k}}
-M(\vec{k}^2)\right]_{\al\ba}}{2\sqrt{\vec{k}^2+M(\vec{k}^2)^2}},\;\;\;\;
M(\vec{k}^2)=\frac{B_m(\vec{k}^2)}{A_s(\vec{k}^2)}.
\ee
The static quark propagator can thus be written in terms of a single mass 
function, $M$.  This is analogous to the case for the static gluon 
propagator.  Expanding $\tilde{F}$ with \eq{eq:tf0}, the equations for 
$A_s$ and $B_m$ can be written as
\bea
A_s(\vec{k}^2)&=&1+\frac{1}{2}\frac{{\cal C}}{\sqrt{\vec{k}^2
+M(\vec{k}^2)^2}}+\frac{1}{2}\int\frac{\dk{\vec{\w}}\s{\vec{k}}{\vec{\w}}
F(\vec{k}-\vec{\w})}{\vec{k}^2\sqrt{\vec{w}^2+M(\vec{w}^2)^2}},\nonumber\\
B_m(\vec{k}^2)&=&m+\frac{1}{2}\frac{{\cal C}M(\vec{k}^2)}{\sqrt{\vec{k}^2
+M(\vec{k}^2)^2}}+\frac{1}{2}\int\frac{\dk{\vec{\w}}M(\vec{\w}^2)
F(\vec{k}-\vec{\w})}{\sqrt{\vec{w}^2+M(\vec{w}^2)^2}}.
\label{eq:dqgap0}
\eea
Clearly the functions $A_s$, $B_m$ (and hence the full quark propagator) are 
dependent on ${\cal C}$ and involve potentially infrared divergent 
integrals, just as for the gluon.  Combining the above equations, we see 
that the ${\cal C}$-dependence of $M$ cancels:
\be
M(\vec{k}^2)=m+\frac{1}{2}\int
\frac{\dk{\vec{\w}}F(\vec{k}-\vec{\w})}{\sqrt{\vec{w}^2+M(\vec{w}^2)^2}}
\left[M(\vec{\w}^2)-\frac{\s{\vec{k}}{\vec{\w}}}{\vec{k}^2}M(\vec{k}^2)\right]
\label{eq:qgap0}
\ee
and one again sees that although $F$ may be strongly infrared divergent, 
$M$ can still be infrared finite.  Equation~(\ref{eq:qgap0}) is the quark 
gap equation and moreover, is well-known from the literature as the 
Adler-Davis truncation \cite{Adler:1984ri}, where it was derived using the 
canonical Hamiltonian approach.  A more advanced version of the quark gap 
equation to self-consistently include the spatial quark gluon vertex has 
been recently studied in the canonical approach \cite{Pak:2011wu}.  The 
equivalence of \eq{eq:qgap0} to the previously derived gap equation for the 
static propagator in the canonical approach \cite{Adler:1984ri} is again a 
result of this study.  The similarity of the quark gap equation, 
\eq{eq:qgap0}, to its gluonic counterpart, \eq{eq:ggap0}, is striking 
(and this is why we take the liberty in referring to \eq{eq:ggap0} as the 
gluon gap equation) and arises primarily from the truncation to include only 
those terms originating from the Coulomb kernel, which itself involves both 
the quark and gluon contributions to the color charge on equal footing.

It is possible to make a connection between the (full) quark propagator and 
the leading order heavy quark propagator in Coulomb gauge 
\cite{Popovici:2010mb} (see also \cite{Popovici:2010ph,Popovici:2011yz}).  
This is based on a spin-decomposition and we follow the spirit of 
Ref.~\cite{Eichten:1980mw}.  Let us introduce the spin-projection operators
\be
P_\pm=\frac{1}{2}\left(\openone\pm\ga^0\right),\;\;\;\;P_++P_-=\openone,\;\;
P_+P_-=0,\;\;P_\pm^2=P_\pm.
\ee
These projectors furnish the following identities:
\be
P_+\ga^0P_+=P_+P_+,\;\;P_-\ga^0P_-=-P_-P_-,\;\;
P_+\ga^0P_-=P_+\ga^iP_+=P_-\ga^iP_-=0
\ee
which allow us to write the quark propagator as
\bea
W_{\ov{q}q\al\ba}(k)&=&
\left[\left(P_++P_-\right)W_{\ov{q}q}(k)\left(P_++P_-\right)\right]_{\al\ba}
\nonumber\\
&=&\frac{(-\imath)}{\left[k_0^2-\vec{k}^2A_s^2(\vec{k}^2)
-B_m^2(\vec{k}^2)+\imath0_+\right]}\nonumber\\&&\times
\left\{\left[k_0+B_m(\vec{k}^2)\right]P_+P_+
-\left[k_0-B_m(\vec{k}^2)\right]P_-P_-
-\left[P_+\s{\vec{\ga}}{\vec{k}}P_-+P_-\s{\vec{\ga}}{\vec{k}}P_+\right]
A_s(\vec{k}^2)\right\}_{\al\ba}.
\eea
The heavy quark limit for Coulomb gauge (implicitly in the rest frame) can 
be expressed as the limit $|\vec{k}|/m\rightarrow0$.  Considering 
\eq{eq:qgap0}, we can make a leading order estimate for the function $M$ 
(and we shall see that this is confirmed by the numerical results).  Given 
that $F(\vec{k}-\vec{\w})$ in the integral peaks at $\vec{\w}=\vec{k}$, the 
infrared divergence is canceled and leaves
\be
M(\vec{k}^2)\approx m+\#\frac{M(\vec{k}^2)}{\sqrt{\vec{k}^2+M^2(\vec{k}^2)}}
\stackrel{|\vec{k}|\ll m}{\rightarrow}m+\#.
\label{eq:heavy0}
\ee
The functions $A_s$, $B_m$ given by \eq{eq:dqgap0} can also then be estimated:
\bea
A_s(\vec{k}^2)&\stackrel{|\vec{k}|\ll m}{\rightarrow}&1
+{\cal O}(1/m)\nonumber\\
B_m(\vec{k}^2)&\stackrel{|\vec{k}|\ll m}{\rightarrow}&m+\frac{1}{2}{\cal C}
+\frac{1}{2}\int\dk{\vec{\w}}F(\vec{\w}^2)\;\;\;(=B_m)
\label{eq:bm0}
\eea
where in $B_m$, it is recognized that the integral is infrared divergent and 
not suppressed by factors $1/m$.  We demand that $|\vec{k}|A_s\rightarrow0$, 
despite the fact that ${\cal C}\rightarrow\infty$ (in the heavy quark limit, 
$m$ is the largest scale).  The spin-decomposed quark propagator in the 
heavy quark limit is then
\bea
W_{\ov{q}q\al\ba}(k)&\stackrel{|\vec{k}|\ll m}{\rightarrow}&
\frac{(-\imath)}{\left[k_0^2-B_m^2+\imath0_+\right]}
\left\{\left[k_0+B_m\right]P_+P_+-\left[k_0-B_m\right]P_-P_-\right\}_{\al\ba}
\nonumber\\
&=&-\imath\frac{\left[P_+P_+\right]_{\al\ba}}{\left[k_0-B_m+\imath\e\right]}
+\imath\frac{\left[P_-P_-\right]_{\al\ba}}{\left[k_0+B_m-\imath\e\right]}.
\eea
The first component represents a heavy quark propagating forward in time; 
the second, a heavy antiquark propagating backwards in time (and under 
time-reversal is equivalent to the first component).  The first component 
is explicitly that found in Ref.~\cite{Popovici:2010mb} (and where only 
forward propagation is included), showing that the leading loop order \DS 
truncation scheme considered here reduces in the heavy quark limit to the 
scheme considered previously.  Notice that when considering the heavy quark 
limit for the Bethe-Salpeter equation \cite{Popovici:2010mb}, the Faddeev 
equation \cite{Popovici:2010ph} and the quark four-point \DS equation 
\cite{Popovici:2011yz}, the constant $\cc$ and the infrared divergence of 
the spatial integral over $F$ that occur in $B_m$, \eq{eq:bm0}, cancel 
explicitly when considering color singlet quark configurations.

\section{\label{sec:numres}Truncated \DS equations: numerical analysis}
\setcounter{equation}{0}
Let us now consider the numerical solutions to the gap equations 
(\ref{eq:ggap0},\ref{eq:qgap0}) for the static dressing functions $G$ and 
$M$.  The defining feature of the truncated gap equations is the input 
Coulomb kernel $F\sim1/\vec{k}^4$.  The strong singularity is also the 
defining problem in solving the equations.

\subsection{gluon gap equation}
In order to solve the gluon gap equation, \eq{eq:ggap0}, it proves 
convenient to change integration variables such that the radial integration 
momentum goes through the Coulomb kernel.  The equation thus reads (to make 
the equation more readable, we use subscripts to denote the momentum 
dependence of the functions)
\be
G_k=1+\frac{N_c}{4C_F}\int\frac{\dk{\vec{\w}}}{\sqrt{(\vec{k}-\vec{\w})^2}}
F_\w t_{ji}(\vec{k})t_{ij}(\vec{k}-\vec{\w})
\left[G_{k-\w}^{1/2}-\frac{(\vec{k}-\vec{\w})^2}{\vec{k}^2}
\frac{G_k}{G_{k-\w}^{1/2}}\right].
\ee
The above equation is not renormalized nor is it regularized.  It is 
convenient to use the following notation:
\be
x=\vec{k}^2,\;\;y=\vec{\w}^2,\;\;
\th=(\vec{k}-\vec{\w})^2=x+y-2\sqrt{xy}z,\;\;
\int\dk{\vec{\w}}\rightarrow
\frac{2}{(4\pi)^2}\int_\e^\La dy\sqrt{y}\int_{-1}^1dz
\ee
where both an ultraviolet (UV), $\La\rightarrow\infty$, and an infrared 
(IR), $\e\rightarrow0$, spatial momentum cutoff have been introduced to 
regularize the integrals.  Inserting the first form of \eq{eq:tf0} for $F$, 
the equation reads
\be
G_x=1+\si_c\ga\int_\e^\La\frac{dy}{y}
\int_{-1}^1\frac{dz}{\sqrt{y\th}}K(x,y;z)
\left[G_\th^{1/2}-\frac{\th}{x}\frac{G_x}{G_\th^{1/2}}\right],
\label{eq:gdse1}
\ee
where
\be
\ga=\frac{N_c^2}{\pi(N_c^2-1)},\;\;K(x,y;z)=1-\frac{y(1-z^2)}{2\th}.
\ee
One immediately sees the effect of the strong IR singularity present in the 
kernel: as $y\rightarrow0$, there must exist some cancellation such that the 
solution is independent of the IR-cutoff ($\e$).  In effect, the combination 
of terms in the last bracket must cancel to leading order in $y$, such that 
the angular integral vanishes as $y\rightarrow0$ 
(and for all values of $x$).  Let us therefore consider the angular integral 
in some more detail:
\be
I_z(x,y)=\int_{-1}^1\frac{dz}{\sqrt{y\th}}K(x,y;z)
\left[G_\th^{1/2}-\frac{\th}{x}\frac{G_x}{G_\th^{1/2}}\right].
\ee
For large values of $x$ and vanishing $y$, $\th=x+\co(y)$ and 
$G_\th\rightarrow G_x+\co(y)$.  Then, it is obvious that
\be
I_z(x,y)\sim\int_{-1}^1\frac{dz}{\sqrt{yx}}
\left[1-\frac{y(1-z^2)}{2x}\right]\left[G_x^{1/2}\co(y)\right]
\ee
which would then automatically lead to an overall IR convergent radial 
integral.  The case where both $x$ and $y$ are small is slightly more 
complicated.  In the process of regularization, an additional (mass) scale 
is introduced so let us assume that to leading order in the IR,
\be
G_x\stackrel{x\rightarrow0}{\rightarrow}G_0\left(\frac{x}{\ka_0}\right)^\al,
\ee
where $\ka_0$ has dimension $[mass]^2$ and $G_0$ is some constant.  The 
angular integral would then read
\be
I_z(x,y)=G_0^{1/2}\int_{-1}^1\frac{dz}{\sqrt{y\th}}K(x,y;z)
\left(\frac{\th}{\ka_0}\right)^{\al/2}
\left[1-\left(\frac{x}{\th}\right)^{\al-1}+\co(y)\right]
\ee
which can only have the necessary cancellation (given that $x$ may be of 
the same order as $y$, but is still unfixed) for $\al=1$.  Thus, the 
cancellation of the IR singularity leads us to consider a solution of the 
form
\be
G_x=\frac{x}{x+\ka_x}
\ee
where for low x, $\ka_x$ should be a constant.  Indeed, in 
Ref.~\cite{Szczepaniak:2001rg} where the gluon gap equation was solved with 
an infrared enhanced (but not $1/\vec{k}^4$) Coulomb kernel, such a 
solution with a constant $\ka_x$ was shown to be a good approximation to 
the full solution.  Note that for such a solution, the static gluon 
propagator ($\sim G_x^{1/2}/\sqrt{x}$) has a constant asymptotic value in 
the IR, characteristic of a massive solution and contradicting the analysis 
of \cite{Zwanziger:1991gz} (see also, Ref.~\cite{hep-th/0605115}), i.e., the 
propagator does not have the Gribov form \cite{Gribov:1977wm}.

The gluon gap equation, \eq{eq:gdse1} can be recast as an equation for 
$\ka_x$:
\be
\ka_x=\si_c\ga\int_\e^\La\frac{dy}{y}\int_{-1}^1 dzK(x,y;z)
\frac{\left[\th-x+\ka_\th-\ka_x\right]}{\sqrt{y}\sqrt{\th+\ka_\th}}.
\label{eq:gdse2}
\ee
Let us discuss the UV behavior of $\ka_x$.  Assuming that for finite, but 
large $x$, $x\gg\ka_x$, then
\be
\ka_x\approx\si_c\ga\int_\e^\La\frac{dy}{y}\int_{-1}^1 dzK(x,y;z)
\frac{\left[\th-x\right]}{\sqrt{y\th}}.
\ee
For $y\gg x$ (the upper limit of the radial integral), we would then have
\be
\ka_x\sim\si_c\ga\int^\La\frac{dy}{y}\int_{-1}^1 dz\frac{1}{2}(1+z^2)\sim 
\frac{4}{3}\si_c\ga\ln(\La)
\ee
which after balancing dimensions leads us to the leading UV behavior for 
$\ka_x$:
\be
\ka_x\sim-\frac{4}{3}\si_c\ga\ln\left(\frac{x}{\La}\right).
\label{eq:b1pred0}
\ee
Thus, although there is no perturbative content in the Coulomb kernel 
($\sim1/\vec{k}^4$), there is still a logarithmic UV-divergent contribution 
to the static gluon propagator.  Moreover, the factor $4/3$, arising from 
the angular integral is characteristic to the ghost self-energy (see e.g., 
Ref.~\cite{Watson:2010cn}).  It appears rather ironic that the function 
$\ka_x$, which was introduced to cancel the leading IR singularity, is 
connected to the UV-cutoff, but in retrospect this is quite natural: the IR 
scale appears during regularization (breaking the scale invariance of the 
original theory).  Further, the divergent term originates in the 
$\G_{AA}$-equation which, were one to have the perturbative form for $F$ 
($\sim1/\vec{k}^2$), gives rise to a quadratic UV-divergence that is 
canceled by the gluon loop.  Here, the quadratic divergence is reduced to a 
logarithmic divergence.  One sees therefore that in Coulomb gauge, the 
nonperturbative UV-limit is not necessarily the same thing as the 
perturbative limit.

Equation~(\ref{eq:gdse2}) is solved by iteration, to give the unrenormalized 
function $\ka_x$ in terms of the UV-cutoff $\La$.  In order to numerically 
cope with the infrared divergence, the equation is rewritten in the form
\be
\ka_x(\La)=\frac{I_1(x;\e,\La)}{1+I_2(x;\e,\La)}
\label{eq:gdse3}
\ee
where
\bea
I_1(x;\e,\La)&=&\si_c\ga\int_\e^\La\frac{dy}{y}\int_{-1}^1 dzK(x,y;z)
\frac{\left[\th-x+\ka_\th\right]}{\sqrt{y}\sqrt{\th+\ka_\th}},\nonumber\\
I_2(x;\e,\La)&=&\si_c\ga\int_\e^\La\frac{dy}{y}\int_{-1}^1 dzK(x,y;z)
\frac{1}{\sqrt{y}\sqrt{\th+\ka_\th}}.
\eea
The reason for using this form is that during iteration, where $\ka$ within 
the integral is not yet the solution, the infrared cancellation required for 
\eq{eq:gdse2} may not happen in practice and the error is amplified by the 
latent infrared singularity: the iteration procedure would thus be 
unstable.  Both $I_1$ and $I_2$ diverge as $\e\rightarrow0$ (the divergence 
is $\sim1/\sqrt{\e}$ in both cases) such that numerically their difference 
will be inherently prone to a loss of fidelity and the residues of the 
singularities are not known exactly enough to compensate.  The ratio of the 
two integrals is better behaved such that with the form \eq{eq:gdse3}, the 
iteration is stable (but only slowly convergent).  In the end, both forms of 
the equation are satisfied.

\begin{figure}[t]
\vspace{0.8cm}
\includegraphics[width=0.5\linewidth]{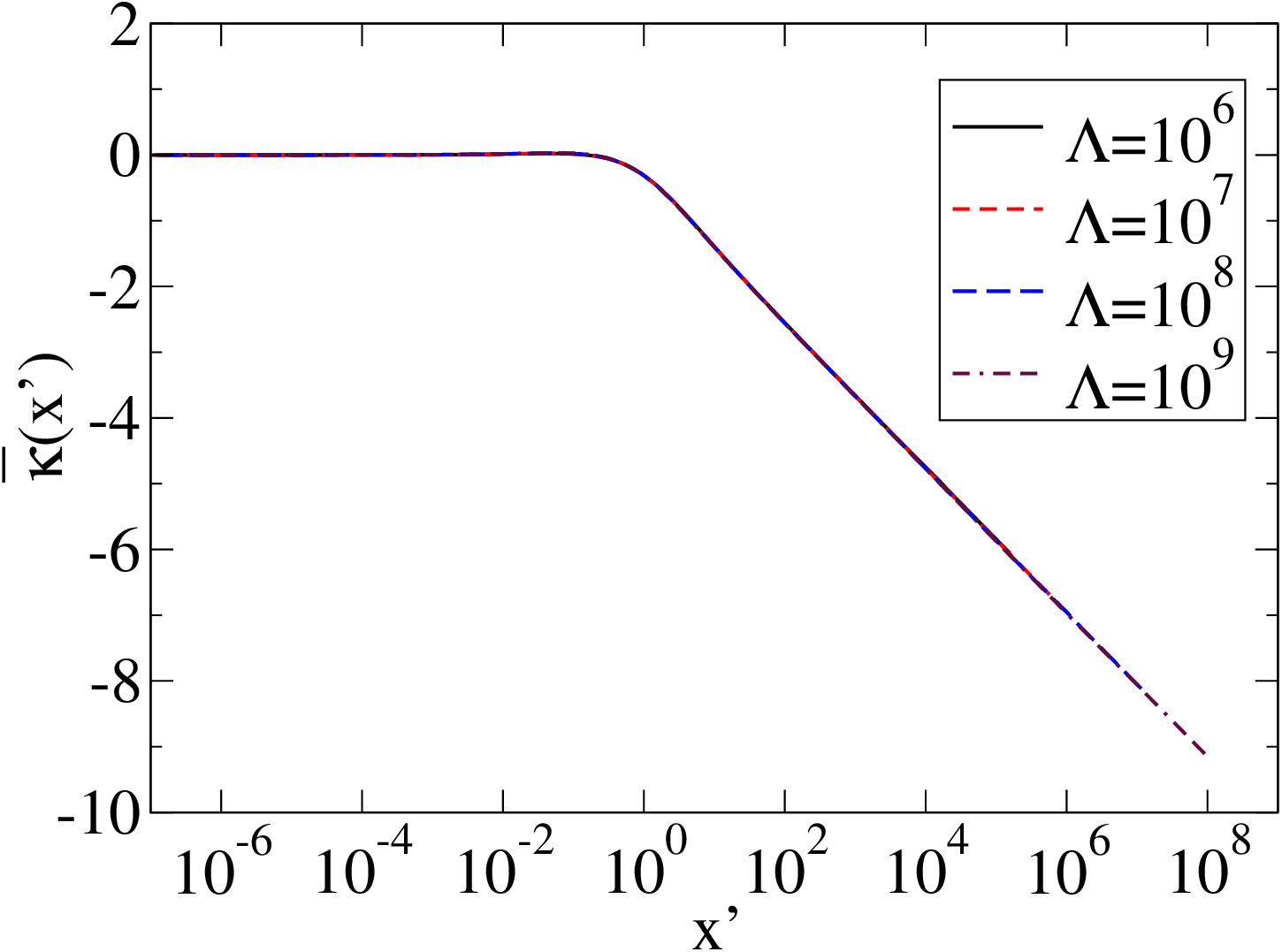}
\vspace{0.3cm}
\caption{\label{fig:hap1}Plot of $\ov{\ka}(x')$ for various values of the 
UV-cutoff, $\La$.  All dimensionful quantities are in units of $\si_c$.  
See text for details.}
\end{figure}

As input, we take the values $N_c=3$ and $\si_c=1$ (all dimensionful 
quantities are thus expressed in units of $\si_c$ for now).  The radial and 
angular integrals are performed using (standard) Gaussian quadrature 
grids.  In the integrand, $\ka_\th$ is evaluated using either cubic spline 
interpolation derived from the grid of $x$-values or, for asymptotic 
arguments, extrapolation based on the following formulae:
\be
\ka_x=\left\{\begin{array}{cc}a_0+a_1x+a_2x^2,&\mbox{small $x$}\\
b_0+b_1\ln(x),&\mbox{large $x$}\end{array}\right..
\label{eq:asymp0}
\ee
In practice, the fit to the UV asymptotic form is performed for 
$x\in[\La/10^3,\La/10]$ in order to avoid complications arising when 
$x\sim\La$.  It is found that for $\e\leq10^{-6}$, $\ka_x$ becomes 
independent of $\e$, as it should.  $\ka_x$ is plotted in 
Fig.~\ref{fig:hap0} for various values of $\La$.  It is seen that in the 
IR, $\ka_x$ goes to a constant value, whereas in the UV, it decreases 
logarithmically, confirming the previous analysis.  The extracted values 
for the asymptotic coefficients $a_0$ (the infrared constant value of 
$\ka_x$) and $b_1$ (the slope of the UV-logarithmic divergence) are given 
in Table~\ref{tab:hap0}.  The analytic value for $b_1$ according to 
\eq{eq:b1pred0} is $-0.477$ with the current input and one sees that this is 
numerically verified (to within a reasonable numerical precision of $<1\%$, 
recalling that the integrals themselves involve $\co(1/\sqrt{\e})$ 
contributions).  Also given in Table~\ref{tab:hap0} are values for an 
estimate for $a_0$ ($\ov{a}_0$).  This is generated by assuming that 
$\ka_x=\ov{a}_0$ is constant and then considering \eq{eq:gdse2} for $x=0$:
\begin{figure}[t]
\vspace{0.8cm}
\includegraphics[width=0.5\linewidth]{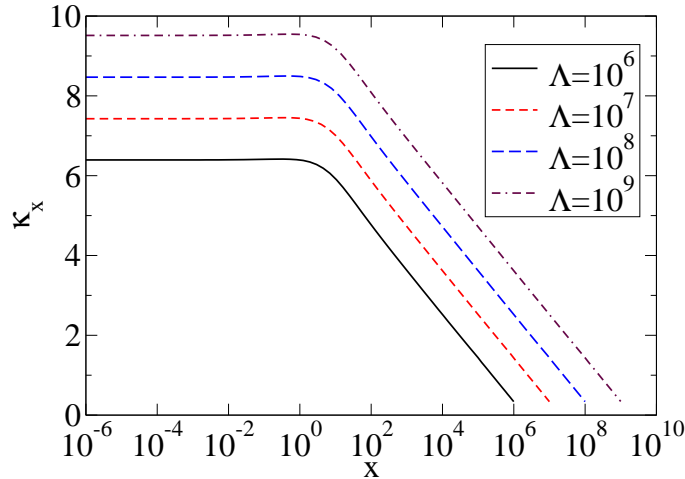}
\caption{\label{fig:hap0}Plot of $\ka_x$ for various values of the 
UV-cutoff, $\La$.  All dimensionful quantities are in units of $\si_c$.  
See text for details.}
\end{figure}
\begin{table}
\begin{tabular}{cccc}\hline\hline
$\La$&$a_0$&$\ov{a}_0$&$b_1$\\\hline
$10^6$&$6.39$&$6.37$&$-0.475$\\
$10^7$&$7.43$&$7.40$&$-0.475$\\
$10^8$&$8.47$&$8.44$&$-0.474$\\
$10^9$&$9.51$&$9.48$&$-0.474$\\\hline\hline
\end{tabular}
\caption{\label{tab:hap0}Numerical values for fitted coefficients for 
various values of $\La$.  All dimensionful quantities are in units of 
$\si_c$.  See text for details.}
\end{table}
\be
\ov{a}_0=\frac{4}{3}\si_c\ga\int_\e^\La\frac{dy}{\sqrt{y}\sqrt{y+\ov{a}_0}}
\approx\frac{4}{3}\si_c\ga\ln{\left(\frac{4\La}{\ov{a}_0}\right)}.
\ee
This estimate is useful for understanding the $\La$-dependence of the full 
solution $\ka_x$.  $\ov{a}_0$ scales \emph{almost} logarithmically with 
$\La$, but not quite; the same is true for $\ka_x$.  Empirically, we find 
that
\be
\ov{\ka}(x')=\ka(x=x'a_0(\La);\La)-a_0(\La)
\label{eq:scale0}
\ee
is independent of $\La$, shown in Fig.~\ref{fig:hap1}.

The appearance of the scaled argument $x'=x/a_0(\La)$ in \eq{eq:scale0} may 
at first seem somewhat arbitrary.  We have seen that the presence of the 
$\si_c/\vec{k}^4$ term in the Coulomb kernel leads to the existence of an 
infrared mass scale $a_0$ that is generated by the regularization scale 
$\La$.  The process of renormalization is tantamount to choosing a reference 
scale, relative to which all other quantities are expressed.  Importantly, 
both the function $\ka_x$ and the string tension $\si_c$ have dimension 
$[mass]^2$ and must therefore be measured in the appropriate units.   
Choosing the (nonperturbatively generated) scale $a_0(\La)$ as our reference 
scale (thereby implicitly choosing the renormalization subtraction point 
$\mu=0$ as will be seen below), we can rewrite the original equation for 
$\ka_x$, \eq{eq:gdse2}, in terms of rescaled quantities.  Restoring the 
putative string tension, $\si_c$, and denoting all scaled quantities with 
a prime:
\be
x'=\frac{x}{a_0(\La)},\;\;\si'_c=\frac{\si_c}{a_0(\La)},\;\;
\ka'=\frac{\ka}{a_0(\La)},\ldots
\ee
(similarly for $y$, $\th$, $\e$ and $\La$ itself), \eq{eq:gdse2} can be 
rewritten as
\be
\ka'(x';\si'_c,\La')
=\si'_c\ga\int_{\e'}^{\La'}\frac{dy'}{y'}\int_{-1}^1dzK(x',y';z)
\frac{\th'-x'+\ka'(\th';\si'_c,\La')-\ka'(x';\si'_c,\La')}
{\sqrt{y'}\sqrt{\th'+\ka'(\th';\si'_c,\La')}}.
\ee
Notice that the equation for $\ka'$ does not explicitly include reference 
to $a_0(\La)$: the $a_0(\La)$-dependence resides in the condition
\be
a_0(\La)=\ka(x=0;\si_c,\La)
=a_0(\La)\ka'(x=x'a_0(\La)=0;\si_c=\si'_ca_0(\La),\La=\La'a_0(\La))
\ee
(technically, $a_0$ is also dependent on $\si_c$, but for notational 
convenience we shall drop the label) such that one may regard $\ka'$ as a 
function of primed quantities and where $\ka'(x'=0;\si'_c,\La')=1$.  
Subtracting at $x'=0$ and defining
\be
\Delta\ka'(x',0;\si'_c)=\ka'(x';\si'_c,\La')-\ka'(0;\si'_c,\La')
=\ka'(x';\si'_c,\La')-1
\ee
then we arrive at a closed expression for $\Delta\ka'(x',0;\si'_c)$:
\begin{align}
\Delta\ka'(x',0;\si'_c)=\si'_c\ga\int_{\e'}^{\La'}\frac{dy'}{y'}&
\left\{\int_{-1}^1 dzK(x',y';z)
\frac{\th'-x'+\Delta\ka'(\th',0;\si'_c)-\Delta\ka'(x',0;\si'_c)}
{\sqrt{y'}\sqrt{1+\th'+\Delta\ka'(\th',0;\si'_c)}}
\right.\nonumber\\&\left.
-\frac{4}{3}\frac{y'+\Delta\ka'(y',0;\si'_c)}
{\sqrt{y'}\sqrt{1+y'+\Delta\ka'(y',0;\si'_c)}}\right\}.
\end{align}
$\Delta\ka'(x',0;\si'_c)$ is explicitly independent of $\La'$, just as a 
renormalized quantity should be.  It is however dependent on $\si'_c$, the 
scaled string tension.  We see that, rather than in the unrenormalized case 
where everything is expressed in units of the string tension, in the 
renormalized case, everything is expressed in units of an implicit 
nonperturbatively generated scale.

\begin{figure}[t]
\vspace{0.8cm}
\includegraphics[width=0.5\linewidth]{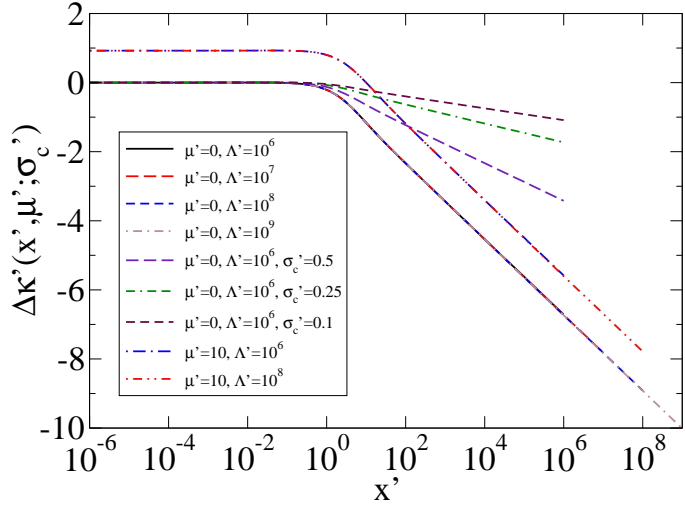}
\vspace{0.3cm}
\caption{\label{fig:tap0}$\Delta\ka'(x',\mu';\si'_c)$ for various values 
of $\La'$, $\si'_c$ ($=1$ unless otherwise stated) and two different 
$\mu'$.  See text for details.}
\end{figure}

The subtraction at $x=x'=0$ is not the only possibility.  Subtraction at 
$x'=\mu'$ is also possible and in this case the scale 
$a(\La)=\ka(x=\mu'a(\La);\si_c,\La)$ is chosen as a reference, such that
\be
\Delta\ka'(x',\mu';\si'_c)=\ka'(x';\si'_c,\La')-\ka'(\mu';\si'_c,\La')
=\ka'(x';\si'_c,\La')-1
\ee
and which obeys
\be
\Delta\ka'(x',\mu';\si'_c)=\si'_c\ga\int_{\e'}^{\La'}\frac{dy'}{y'}
\left\{\int_{-1}^1 dzK(x',y';z)\frac{\th'-x'+\Delta\ka'(\th',\mu';\si'_c)
-\Delta\ka'(x',\mu';\si'_c)}
{\sqrt{y'}\sqrt{1+\th'+\Delta\ka'(\th',\mu';\si'_c)}}
-(x'\rightarrow\mu')\right\}.
\ee
The slope of the curves in the UV is proportional to $\si'_c$ and is the 
same for each value of $\mu'$ -- the \emph{scaled} string tension $\si'_c$ 
is independent of the renormalization point, although the scaling factor, 
$a(\La)$, is dependent on the choice of subtraction point $\mu'$.  This is 
nothing other than the statement that $\si'_c$ is the external input to the 
renormalized equation.  $\Delta\ka'(x',\mu';\si'_c)$ is plotted in 
Fig.~\ref{fig:tap0} for various values of $\La'$, $\si'_c$ and two 
different $\mu'$.

\begin{figure}[t]
\vspace{0.8cm}
\includegraphics[width=0.5\linewidth]{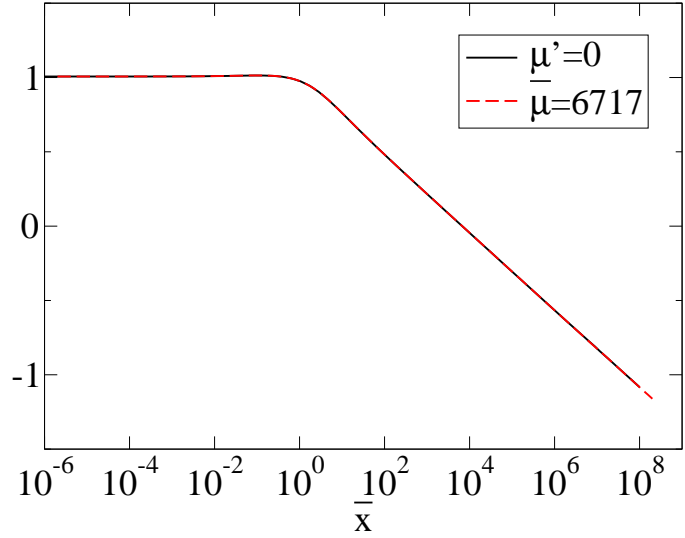}
\vspace{0.3cm}
\caption{\label{fig:tap1}Renormalized functions $f(\ov{x})$ (labeled 
as $\mu'=0$) and $\Delta\ov{\ka}(\ov{x},\ov{\mu};\ov{\si}_c)$ (labeled 
$\ov{\mu}=6717$) plotted versus $\ov{x}$.  See text for details.}
\end{figure}

We notice that in Fig.~\ref{fig:tap0}, the curves for different $\mu'$ 
follow the same pattern as for the unrenormalized case, Fig.~\ref{fig:hap0} 
with different values of $\La$.  The following question arises: given 
$\Delta\ka'(x',\mu';\si'_c)$, can we derive the renormalized function for a 
different subtraction point, $\ov{\mu}$, and what are the associated scale 
factors?  To answer the question, let us take a step back to the original 
unrenormalized function $\ka(x;\si_c,\La)$ and define
\bea
a(\La)&=&\ka(x=\mu'a(\La);\si_c=\si'_ca(\La),\La=\La'a(\La)),\nonumber\\
b(\La)&=&\ka(x=\ov{\mu}b(\La);\si_c=\ov{\si}_cb(\La),\La=\ov{\La}b(\La)).
\eea
So, knowing the unrenormalized function $\ka(x;\si_c,\La)$, $a(\La)$ and 
$b(\La)$ can be derived for the chosen $\mu'$ and $\ov{\mu}$.  Further 
defining the scaled functions as before:
\bea
\ka(x;\si_c,\La)&=&a(\La)\ka'(x=x'a(\La);\si_c=\si'_ca(\La),\La=\La'a(\La)),
\nonumber\\
&=&b(\La)\ov{\ka}(x=\ov{x}b(\La);\si_c=\ov{\si}_cb(\La),\La=\ov{\La}b(\La)),
\label{eq:scale1}
\eea
such that when written in terms of the appropriately scaled variables
\be
\ka'(x'=\mu';\si'_c,\La')=\ov{\ka}(\ov{x}=\ov{\mu};\ov{\si}_c,\ov{\La})=1.
\ee
The renormalized ($\La$-independent) functions are then
\bea
\Delta\ka'(x',\mu';\si'_c)&=&
\ka'(x=x'a(\La);\si_c=\si'_ca(\La),\La=\La'a(\La))-1,\nonumber\\
\Delta\ov{\ka}(\ov{x},\ov{\mu};\ov{\si}_c)&=&
\ov{\ka}(x=\ov{x}b(\La);\si_c=\ov{\si}_cb(\La),\La=\ov{\La}b(\La))-1.
\eea
Importantly, both the above functions can be related to the original 
unrenormalized function $\ka$ and the original variables through 
\eq{eq:scale1}.  One thus sees that
\be
\Delta\ov{\ka}(\ov{x},\ov{\mu};\ov{\si}_c)+1
=\frac{a(\La)}{b(\La)}\left[
\Delta\ka'\left(x'=\ov{x}\frac{b(\La)}{a(\La)},\mu';
\si'_c=\ov{\si}_c\frac{b(\La)}{a(\La)}\right)+1\right].
\ee
So, one can indeed derive the renormalized function for subtraction point 
$\ov{\mu}$ in terms of that subtracted at $\mu'$ and there is a single 
scale factor given by the ratio
\be
Z(\ov{\mu},\mu';\si_c,\La)=\frac{a(\La)}{b(\La)}
=\frac{\ka(x=\mu'a(\La);\si_c,\La)}{\ka(x=\ov{\mu}b(\La);\si_c,\La)}.
\ee
To test this, we take the unrenormalized function with $\si_c=1$, $\La=10^8$ 
(plotted in Fig.~\ref{fig:hap0}) and consider the case $\mu'=0$ to give 
$a(\La)=8.4679$ and $\si'_c=1/a(\La)$ (implicitly in units of $\si_c$).  
Considering $\ov{\mu}=6.7171\times10^3$, the ratio 
$Z\equiv a(\La)/b(\La)=2.0067$, with $b(\La)=4.2198$ and 
$\ov{\si}_c=1/b(\La)$.  The functions
\be
f(\ov{x})\equiv 
Z\left[\Delta\ka'(x'=\ov{x}/Z,\mu'=0;\si'_c=\ov{\si}_c/Z)+1\right]-1
\ee
and $\Delta\ov{\ka}(\ov{x},\ov{\mu};\ov{\si}_c)$ are plotted in 
Fig.~\ref{fig:tap1} and one sees that they are identical (as they should be).

Returning to the dressing function $G_x$, we see that
\be
G_x=G(x;\si_c,\La)=\frac{x}{x+\ka(x;\si_c,\La)}
=\frac{x'}{x'+1+\Delta\ka'(x',\mu';\si'_c)}=G'(x',\mu';\si'_c).
\ee
When all dimensionful quantities are expressed in terms of the appropriate 
units (i.e., scaled by the nonperturbatively generated scale defined by the 
subtraction point $\mu'$), the dressing function $G$ is automatically 
independent of the UV-cutoff without modification and with a `mass function' 
$1+\Delta\ka'(x',\mu';\si'_c)$ in these units.  (The scaling factor 
$Z(\ov{\mu},\mu';\si_c,\La)$, defined in terms of the unrenormalized 
solution $\ka(x;\si_c,\La)$, tells us how the physical scale varies with 
different renormalization points.)  In other words, the gluon dressing 
function requires no renormalization factor once the physical scale has 
been set.  This has the implication that the nonperturbatively generated 
gluon mass would be an observable under the present truncation and with 
the $1/\vec{q}^4$ interaction, despite the naive appearance of a logarithmic 
UV-divergence.

\subsection{quark gap equation}
As in the case for the gluon gap equation, it is convenient to change the 
momentum routing in the quark gap equation, \eq{eq:qgap0}, such that the 
radial integration momentum goes through the Coulomb kernel.  The 
corresponding equation is then
\be
M_k=m+4\pi\si_c\int
\frac{\dk{\vec{\w}}}{[\vec{\w}^2+\xi]^2
\left[(\vec{k}-\vec{w})^2+M_{k-\w}^2\right]^{1/2}}
\left[M_{k-\w}-\frac{\vec{k}\cdot(\vec{k}-\vec{\w})}{\vec{k}^2}M_k\right]
\label{eq:qgap3}
\ee
where we have inserted the second form for the Coulomb kernel, \eq{eq:tf0}: 
an infrared regularized form for the interaction with $\xi$ having dimension 
$[mass]^2$ and playing the role of a fictitious mass scale and for which the 
limit $\xi\rightarrow0$ will be studied.  (This is the infrared 
regularization method used, for example, in Ref.~\cite{Alkofer:2005ug}.)  As 
has been emphasized, one can see the obvious similarities between the gluon 
and quark gap equations.  In fact, there are some subtle differences arising 
from the specific forms for the integral kernel: in the quark case, the 
infrared singularities are somewhat more difficult to overcome than in the 
gluon case; however, the quark equation is explicitly UV-convergent with the 
above interaction.  Notice that we will use the original form for the string 
tension, $\si_c$, without mentioning the renormalization and scaling factors 
arising from the discussion of the gluon.  For the quark case, all results 
may be expressed in units of $\si_c$ directly (as shall be seen) and one may 
use $\si_c$ or $\si'_c$ interchangeably as input.  Using the conventions for 
the variables as for the gluon equation whereby $\vec{k}^2=x$, etc., the 
equation reads
\be
M_x=m+\frac{\si_c}{2\pi}\int_\e^\La\frac{dy\,\sqrt{y}}{[y+\xi]^2}
\int_{-1}^1\frac{dz}{\left[\th+M_\th^2\right]^{1/2}}
\left[M_\th-M_x+\sqrt{\frac{y}{x}}zM_x\right].
\label{eq:qgap2}
\ee
In the above equation, a second possibility for infrared regularization 
emerges: one can set $\xi=0$ and study the infrared cutoff limit 
$\e\rightarrow0$ instead of the infrared mass regularization.  It will be 
seen that both methods give identical results.

The infrared analysis of \eq{eq:qgap2} follows in the same manner as for 
the gluon.  Setting $\xi=0$ for the moment it follows that for the radial 
integral to converge as $y\rightarrow0$, there must be some cancellation 
within the angular integral and we require that
\be
I_z(x,y)=\int_{-1}^1\frac{dz}{\left[\th+M_\th^2\right]^{1/2}}
\left[M_\th-M_x+\sqrt{\frac{y}{x}}zM_x\right]
\ee
vanishes fast enough in this limit, and for all values of $x$.  Using the 
techniques as before, this is clearly achieved if $M_x$ tends to a constant 
in the infrared (and as will be seen numerically).

The UV analysis is best performed perturbatively.  Consider the following 
integral (restoring here the original variables, $\vec{k}$, etc.):
\be
I=m+4\pi\si_cM_0\int\frac{\dk{\vec{\w}}}
{[\vec{\w}^2]^2\left[(\vec{k}-\vec{w})^2+M_0^2\right]^{1/2}}
\frac{\vec{k}\cdot\vec{\w}}{\vec{k}^2}
\ee
which corresponds to the case of \eq{eq:qgap3} where $\xi=0$ and where the 
function $M$ within the integrand has been replaced by a constant, $M_0$ 
(i.e., the above integral represents a first iteration of the full gap 
equation).  The above integral can be performed using dimensional 
regularization (see, for example, Refs.~\cite{Watson:2007mz,Popovici:2008ty} 
for a discussion of such integrals) and the result for $\vec{k}^2\gg M_0^2$ 
is
\be
I\stackrel{\vec{k}^2\gg M_0^2}{=}m+\frac{\si_cM_0}{\pi\vec{k}^2}
+\co(1/\vec{k}^4).
\ee
Importantly, the integral is explicitly UV-convergent (for all external 
momenta, $\vec{k}^2$) and this means that the solution to the quark gap 
equation requires no renormalization, at least in the absence of the 
perturbative interaction $\sim1/\vec{k}^2$.  Unlike the gluon, the 
nonperturbatively generated dynamical scale thus plays no role in setting 
the physical units.  In the full equation, \eq{eq:qgap3}, one would expect 
that the mass function vanishes in the UV like $1/\vec{k}^2$ when the bare 
quark mass $m\neq0$, since within the integrand, the mass function is 
negligible compared to the large momentum.  In the chiral case, $m=0$, one 
would expect that the mass function vanishes faster than $1/\vec{k}^2$.

Having discussed the asymptotic behavior, let us now turn to the numerical 
solution of \eq{eq:qgap2}.  In the presence of the infrared singular 
integrals, it is again necessary to modify the equation for numerical 
use.  To iterate, we use the form
\be
M_x=\frac{m+I_1(x;\e,\xi,\La)}{1+I_2(x;\e,\xi,\La)}
\label{eq:qgap4}
\ee
where
\bea
I_1(x;\e,\xi,\La)&=&\frac{\si_c}{2\pi}\int_\e^\La\frac{dy\sqrt{y}}{[y+\xi]^2}
\int_{-1}^1\frac{dz\,M_\th}{\left[\th+M_\th^2\right]^{1/2}}\nonumber\\
I_2(x;\e,\xi,\La)&=&\frac{\si_c}{2\pi}\int_\e^\La\frac{dy\sqrt{y}}{[y+\xi]^2}
\int_{-1}^1\frac{dz}{\left[\th+M_\th^2\right]^{1/2}}
\left[1-\sqrt{\frac{y}{x}}z\right]
\eea
and we will consider two cases for the infrared regularization: 
$\e\rightarrow0$ with $\xi=0$ (infrared cutoff) and $\xi\rightarrow0$ with 
$\e\ll\xi$ (infrared mass).  The iteration procedure for \eq{eq:qgap4} is 
stable over a range of values for $\xi$ and $\e$, but becomes progressively 
more difficult as the regulators are made smaller and where the integrals 
diverge as $1/\sqrt{\e}$ or $1/\sqrt{\xi}$.  The integrals are performed as 
for the gluon, the only difference being the asymptotic formula in the UV, 
where a powerlaw form
\be
M_x\stackrel{x\rightarrow\La}{=}m+cx^{-d}
\ee
is used.  Setting $\si_c=1$, all dimensionful quantities are henceforth 
expressed in units of $\si_c$ (or $\si'_c$ if one uses the renormalized 
scale defined via the massive gluon propagator).  Note that the number of 
colors, $N_c$, has already been absorbed into the definition of the Coulomb 
kernel.  Typically, the solutions to \eq{eq:qgap4} are stable for $\La=10^3$ 
in the chiral case and $\La=10^4$ for the massive case (the only exception 
is the heaviest mass $m=10$, for which $\La=10^5$ is necessary, see later 
for details).

\begin{figure}[t]
\vspace{0.8cm}
\includegraphics[width=0.5\linewidth]{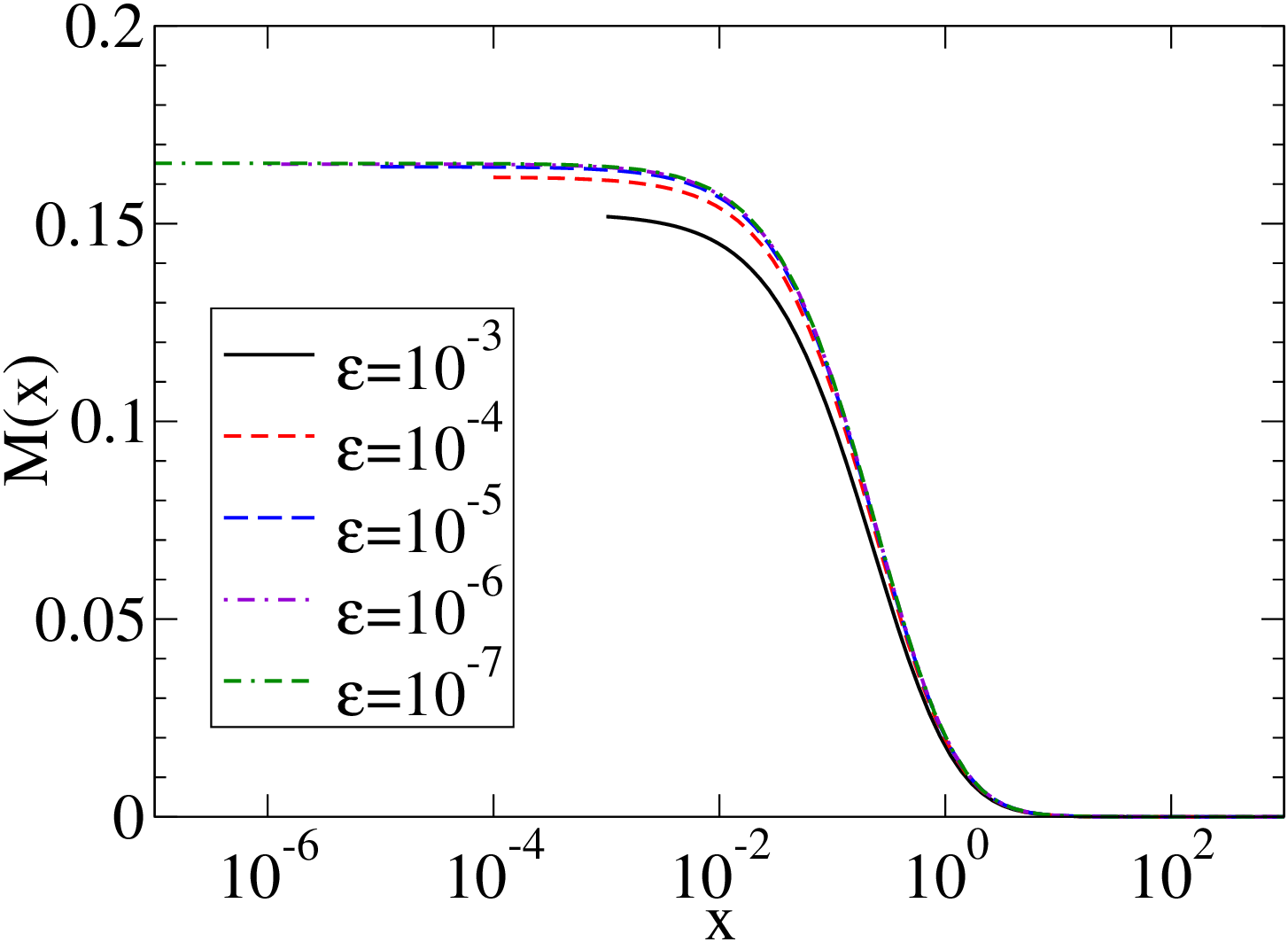}
\vspace{0.3cm}
\caption{\label{fig:aap0}Chiral quark mass function, $M_x$, plotted for 
varying infrared cutoff regulator $\e$.  All dimensionful quantities are 
in appropriate units of the string tension, $\si_c$.  See text for details.}
\end{figure}

Starting with the chiral quark ($m=0$) and with infrared cutoff 
regularization (i.e., studying the limit $\e\rightarrow0$ with $\xi=0$), 
the quark mass function, $M_x$, is plotted in Fig.~\ref{fig:aap0} for 
various values of $\e$.
One sees that as $\e\rightarrow0$, the mass function goes to a constant 
value $M(x=0)=M_0\approx0.165$ (in units of $\sqrt{\si_c}$) in the IR.  The 
solution thus corresponds to a situation whereby chiral symmetry is 
dynamically broken.  Numerically, the behavior in the UV corresponds to a 
powerlaw with exponent $d\approx1.84$ (for the lowest value of $\e$), which 
we shall discuss shortly.

\begin{figure}[t]
\vspace{0.8cm}
\includegraphics[width=0.5\linewidth]{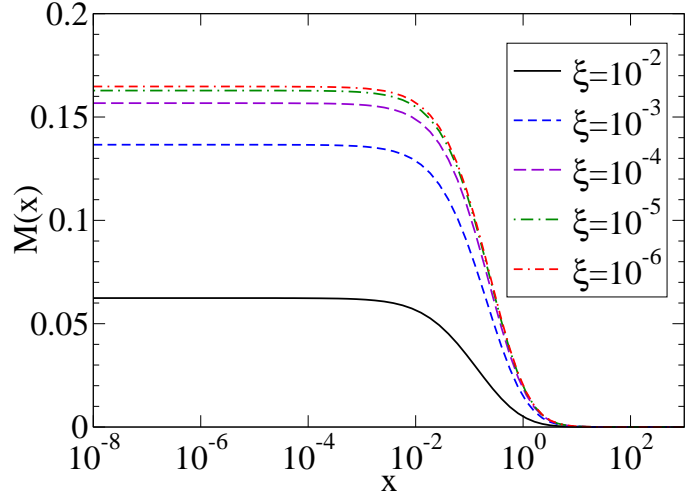}
\vspace{0.3cm}
\caption{\label{fig:bap0}Chiral quark mass function, $M_x$, plotted for 
varying infrared mass regulator $\xi$.  All dimensionful quantities are in 
appropriate units of the string tension, $\si_c$.  See text for details.}
\end{figure}

Repeating the analysis for the chiral quark, but with an infrared mass 
regularization (i.e., studying the limit $\xi\rightarrow0$, with 
$\e<10^{-8}\ll \xi$), the solution is plotted in Fig.~\ref{fig:bap0} for 
various values of $\xi$.
It is seen that as the regularization is removed, the mass function becomes 
identical to that using the infrared cutoff method, showing that the two 
methods agree.  The infrared constant value is the same as before 
$M_0\approx0.165$, although the UV exponent is a little larger: 
$d\approx1.90$.  Both sets of results for the chiral quark may be directly 
compared to those of Ref.~\cite{Alkofer:2005ug} or, allowing for a factor 
$C_F=4/3$ in the definition of $\si_c$, to the results of 
Refs.~\cite{Adler:1984ri,Pak:2011wu}.  Turning to the UV exponent, 
Ref.~\cite{Adler:1984ri} showed that one should expect $d=2$ (in other words 
$M\sim1/\vec{k}^4$ in the UV).  The numerical results here are somewhat 
lower (although within a reasonable numerical precision); however, given 
that the function is vanishing so rapidly, the effect of the UV tail is 
negligible.

\begin{figure}[t]
\vspace{0.8cm}
\includegraphics[width=0.5\linewidth]{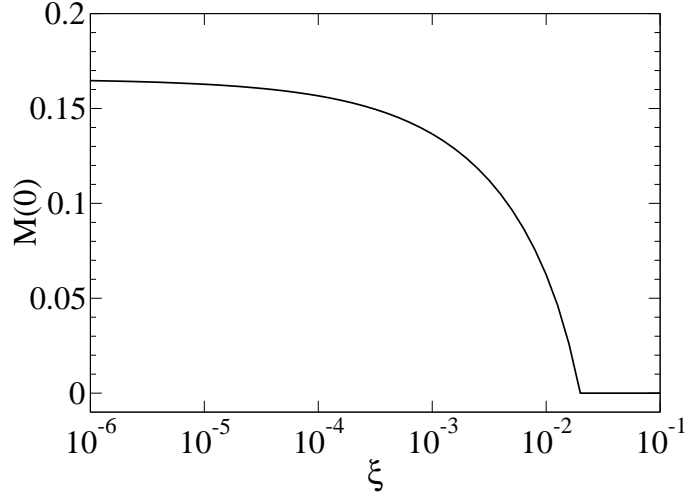}
\vspace{0.3cm}
\caption{\label{fig:cap0}Chiral quark mass function at zero momentum, 
$M(x=0)$, plotted as a function of the infrared mass regulator $\xi$.  All 
dimensionful quantities are in appropriate units of the string tension, 
$\si_c$.  See text for details.}
\end{figure}

The behavior of the mass function for varying infrared mass regulator is 
rather interesting.  In Fig.~\ref{fig:cap0}, the infrared constant mass, 
$M_0$, is plotted as a function of $\xi$.
It is seen that as $\xi\rightarrow0$, $M_0$ tends to its constant, nonzero 
value and indicating that chiral symmetry is dynamically broken in the 
presence of the strongly infrared enhanced Coulomb kernel interaction (and 
which one would naively expect).  However, what is also seen is that for 
large $\xi$, $M_0$ rapidly decreases and vanishes altogether for 
$\xi>0.02$.  As is well known, the interaction $F\sim\si_c/\vec{k}^4$ 
corresponds to a linearly rising potential with a coefficient given by 
$\si_c$.  The explicit expression, appropriate to Coulomb gauge in the heavy 
quark limit and under the truncation to neglect pure Yang-Mills vertices 
reads \cite{Popovici:2010mb} (temporarily reinstating the string tension 
$\si_c$ and all other constants):
\be
V(r)=g^2C_F\int\dk{\vec{\w}}\tilde{F}_\w(1-e^{\imath\vec{\w}\cdot\vec{r}})
=\int\dk{\vec{\w}}\left\{\cc(2\pi)^3\de(\vec{\w})
+\frac{8\pi\si_c}{[\vec{\w}^2+\xi]^2}\right\}
(1-e^{\imath\vec{\w}\cdot\vec{r}})
\label{eq:pot0}
\ee
where $r$ is a length scale.  We shall shortly show that this form is indeed 
the correct expression for the truncation scheme considered here.  The 
$\de$-function term proportional to $\cc$ arises from the charge 
conservation term originating in resolving the temporal zero modes inherent 
to Coulomb gauge.  This term does not contribute to the potential: such a 
term was in fact considered in Ref.~\cite{Adler:1984ri}, to the same 
effect.  With the infrared mass regularization, the integral is
\be
V(r)=\frac{\si_c}{\sqrt{\xi}}\left[1-e^{-r\sqrt{\xi}}\right]
=\left\{\begin{array}{cc}
\si_cr,\,\,&r\sqrt{\xi}\rightarrow0\\
\si_c/\sqrt{\xi},\;\;&r\sqrt{\xi}\rightarrow\infty.
\end{array}\right.
\ee
In the limit $\xi\rightarrow0$ and for finite length scale $r$, $V(r)$, is 
a good approximation to the linearly rising potential, which gets better as 
$\xi$ decreases.  However, as $\xi$ increases (whilst keeping $r$ fixed), 
$V(r)$ flattens to a constant and this constant decreases as $\xi$ 
increases.  The restoration of chiral symmetry for large $\xi$ is now 
rather obvious -- for large $\xi$, the long-range part of the potential is 
no longer linearly-rising, but is constant.  Thus, within the truncation 
scheme here, chiral symmetry is always broken by the existence of a pure 
linearly rising potential, the dynamically generated quark mass being 
measured in units of $\sqrt{\si_c}$.  One cannot speak of a critical string 
tension in this respect (as compared to the concept of a critical 
coupling).  However, the dynamical breaking of chiral symmetry is sensitive 
to the details of the potential when there is a flattening at large range, 
caused in this case by the large value of the infrared mass regulator.

\begin{figure}[t]
\vspace{0.8cm}
\includegraphics[width=0.5\linewidth]{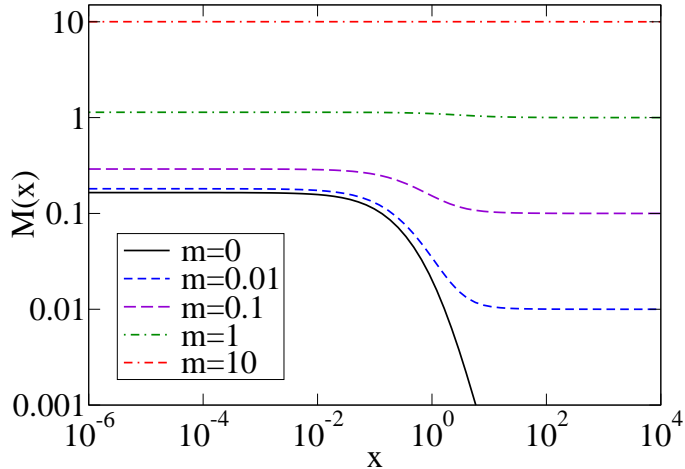}
\vspace{0.3cm}
\caption{\label{fig:eap0}Quark mass function, $M_x$, plotted for various 
bare quark masses.  All dimensionful quantities are in appropriate units of 
the string tension, $\si_c$.  See text for details.}
\end{figure}

\begin{figure}[t]
\vspace{0.8cm}
\includegraphics[width=0.5\linewidth]{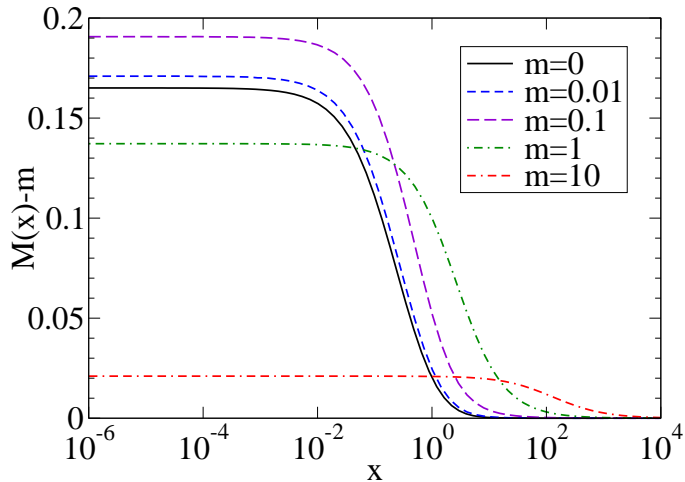}
\vspace{0.3cm}
\caption{\label{fig:eap1}Difference of the quark mass function and the bare 
quark mass, $M_x-m$, plotted for various bare quark masses.  All 
dimensionful quantities are in appropriate units of the string tension, 
$\si_c$.  See text for details.}
\end{figure}

Let us finally discuss the case when the quarks have a nonzero bare mass.  
In the absence of UV-divergences, the bare mass requires no corrections due 
to renormalization.  We again set $\si_c=1$ such that all dimensionful 
quantities are measured in the appropriate unit of the string tension.  
Using the infrared cutoff method, with $\e=10^{-6}$ and $\La=10^4$ (except 
for the case $m=10$, where $\La=10^5$ is used), the solution to the gap 
equation for various quark masses is shown in Fig.~\ref{fig:eap0} (the 
chiral quark is shown for comparison).
One sees that in all cases, the quark mass function is constant in the 
infrared.  In the UV, the exponent of the powerlaw lies in the range 
$d=1.02-1.07$.  This is comparable to the expected value $d=1$ obtained 
earlier from the perturbative analysis, though as for the chiral quarks, 
the UV tail is not particularly important.  In Fig.~\ref{fig:eap1}, we plot 
the deviation of the mass function from the bare quark mass: $M_x-m$.
Two features emerge from this plot.  The first is that as $m$ increases, the 
infrared constant value $M_0-m$ extends further into the UV region (this is 
why $\La$ must be increased for the largest value of $m$, such that the 
solution converges properly).  The second feature is that for increasing 
bare mass, $m$, the difference of the infrared value $M_0-m$ initially 
increases, but then turns over such that for heavy quarks, this difference 
becomes significantly smaller.  In the heavy quark limit, one thus sees 
that $M_x-m$ is suppressed, at least in this truncation.  Moreover, as the 
difference $M_x-m$ vanishes when $m\rightarrow\infty$, then the question of 
the ordering of the limits $m\rightarrow\infty$ and $\La\rightarrow\infty$ 
becomes irrelevant.  One can thus see that the relation \eq{eq:heavy0} is 
valid such that the connection between the truncated quark gap equation 
studied in this work and the heavy quark limit studied analytically in 
Refs.~\cite{Popovici:2010mb,Popovici:2010ph,Popovici:2011yz} (where 
contributions from pure Yang-Mills vertices were neglected) is 
established.  This justifies the earlier use of \eq{eq:pot0}.

\section{\label{sec:conc}Summary, discussion and conclusions}
\setcounter{equation}{0}

In this study, we have considered Coulomb gauge quantum chromodynamics 
within the first order formalism, under a leading order truncation and 
concentrating on the infrared behavior.  In fixing to Coulomb gauge, 
particular attention must be paid to the temporal zero modes of the 
Faddeev-Popov operator.  Within the first order formalism in Coulomb gauge, 
the temporal component of the gauge field can be exactly integrated out, 
leading to the cancellation of the Faddeev-Popov determinant; what remains 
of the gluon sector concerns two transverse field degrees of freedom 
$\vec{A}$ and $\vec{\pi}$.  The resolution of the temporal zero modes 
further leads to the constraint that the total color charge be conserved 
and vanishing.  This constraint was written in Gaussian form, resulting in 
a constant shift in the Coulomb kernel, proportional to 
$\cc\rightarrow\infty$.

Having integrated out the temporal component of the gluon field, the action 
is nonlocal due to the presence of the inverse Faddeev-Popov operator 
occurring in the Coulomb kernel.  In order to derive the \DS equations, a 
leading order truncation was introduced whereby the nonlocal Coulomb kernel 
was replaced by its expectation value in the form of an input Ansatz.  This 
truncation led to the appearance of a set of new (momentum dependent) 
four-point interaction vertices.  These vertices effectively replace the 
dynamical content of the tower of \DS equations and Slavnov-Taylor 
identities involving the temporal, longitudinal and ghost degrees of freedom 
in the local formalism.  The resulting \DS equations were further truncated 
to include (and subsequently study) only those one-loop tadpole terms 
involving the nonperturbative part of the input Ansatz for the Coulomb 
kernel, which was taken to be strongly infrared enhanced ($\sim1/\vec{k}^4$).

It was found that the truncated equation for the mixed gluonic proper 
two-point function, $\G_{\pi A}$, leads immediately to the result that 
this component is trivial.  This in turn meant that in the subsequent 
gluonic \DS equations for $\G_{\pi\pi}$ and $\G_{AA}$, the energy dependence 
could be resolved.  Moreover, the static gluon propagators could be easily 
identified in terms of a single dressing function, $G$.  The proper 
two-point dressing functions were dependent on the constant $\cc$ (arising 
from the charge constraint) and included potentially infrared divergent 
integral contributions (see below for a discussion).  However, combining 
the two \DS equations led to an equation for $G$ alone: \eq{eq:ggap0}.  
This ``gluon gap equation" for the static dressing function $G$ was 
independent of $\cc$ and (as was numerically verified) had a finite 
solution, despite the strongly infrared singular interaction.  An almost 
identical situation was seen for the quark sector: the static propagator 
was given in terms of a single dressing function, $M$, and a gap equation 
arose, \eq{eq:qgap0}, which was independent of $\cc$ and whose solution was 
finite, despite the fact that the \DS equations for the proper dressing 
functions were explicitly dependent on $\cc$ and involved infrared divergent 
integrals.  Importantly, the static gluon and quark gap equations were 
identical in form to their counterparts derived in the canonical Hamiltonian 
approach \cite{Szczepaniak:2001rg,Adler:1984ri}.  This allows for an 
equivalence to be established between the respective truncation schemes and 
approximations employed within the two approaches.

Using spin projectors, it was possible to consider the heavy quark limit and 
it was seen that the (full, not static) quark propagator reduces to the 
known result \cite{Popovici:2010mb} within this truncation.  This is 
important because in the heavy quark limit, the role of the infrared 
divergent integrals is understood: when studying physical, color singlet 
quantities, such infrared divergences cancel whereas for propagators, the 
pole position is shifted to infinity and reflecting the fact that infinite 
energy is required for an unphysical colored object to exist in isolation.  
This in turn explains the constant, $\cc$, stemming from the charge 
constraint: it is merely an additional, fully nonperturbative and constant 
contribution to be added to the infrared divergence.  Given that the input 
Ansatz for the Coulomb kernel was shown to be directly related to the 
instantaneous part of the temporal gluon propagator, which in 
Ref.~\cite{Popovici:2010mb} was shown to be related to the quark-antiquark 
potential, $\cc$ is simply a constant shift of this potential.  In 
Ref.~\cite{Adler:1984ri}, it was demonstrated that an arbitrary, constant 
shift in the potential has no observable consequence (at least as far as 
the chiral quarks studied therein were concerned).  Here, the charge 
constraint arising from the incompleteness of the gauge fixing (the temporal 
zero modes) results in exactly such an unobservable, constant shift 
(considering the limit $\cc\rightarrow\infty$) in the potential, leaving 
physical quantities untouched and naively prohibiting unphysical, color 
charged quantities from existing in isolation (in the sense that infinite 
energy is required to create such an object).

For both the gluonic and quark sectors, it was numerically found that given 
the infrared enhanced input Ansatz, the static propagator dressing functions 
exhibited the presence of a nonperturbatively generated dynamical mass 
scale.  Both functions were nontrivially finite and constant in the 
infrared.  In the case of the gluon, despite the absence of the perturbative 
components, a logarithmic ultraviolet divergence emerged.  However, 
expressing the gluon gap equation in terms of the nonperturbatively 
generated dynamical scale, the dressing function was seen to require no 
renormalization.  This would suggest that the dynamical gluon mass would be 
observable within this leading order truncation insofar as the 
renormalization is concerned (i.e., if one were to ignore the previous 
discussion about the charge constraint and infrared divergences connected 
to the full propagator).  The result is in disagreement with the 
Gribov-Zwanziger confinement scenario 
\cite{Gribov:1977wm,Zwanziger:1995cv,Zwanziger:1998ez} and the results of 
Ref.~\cite{Zwanziger:1991gz}.  This is presumably a shortcoming of the 
leading order truncation scheme utilized in this study.

In the case of the static quark propagator in the chiral limit, the gap 
equation was solved using two infrared regularization procedures and their 
equivalence was numerically demonstrated.  The results are identical to 
previous studies \cite{Alkofer:2005ug,Adler:1984ri,Pak:2011wu}.  The 
nonperturbatively generated mass scale corresponds to dynamical chiral 
symmetry breaking, although one should point out that it is long known that 
quantitatively the resulting chiral condensate is too small 
\cite{Adler:1984ri} (this subject was tackled in Ref.~\cite{Pak:2011wu}).  
Of interest however for the chiral quarks, was the infrared mass 
regularization, where the regularization parameter, $\xi$, was large: in 
this case, the nonperturbatively generated scale decreased as $\xi$ 
increased and eventually disappeared.  Utilizing the connection between the 
input Ansatz for the interaction and the quark-antiquark potential, the 
observed restoration of chiral symmetry could be intuitively explained as 
a flattening of the long-range part of the potential when the interaction 
is regularized.  With heavy quarks, it was explicitly verified numerically 
that in this truncation, the heavy quark limit emerges naturally.

Clearly, despite the fact that we retain only the leading order 
contributions and that the input Ansatz for the Coulomb kernel excludes the 
perturbative content of the theory, the \DS equations of Coulomb gauge in 
the first order formalism represent a powerful tool to study nonperturbative 
quantum chromodynamics.  Equally clearly, one can see that much is missing.  
For example, it is known from the canonical Hamilton approach that: the 
inclusion of the ghost loop (`curvature') in the deep infrared region 
\cite{Feuchter:2004mk,Reinhardt:2004mm} and the triple-gluon vertex in the 
mid-momentum region \cite{Campagnari:2010wc} of the static gluon propagator 
have important roles; also that the spatial quark-gluon vertex is necessary 
to obtain a reasonable estimate for the chiral quark condensate 
\cite{Pak:2011wu}.  Now, in the canonical Hamiltonian approach, the 
truncation scheme is defined by an initial Ansatz for the vacuum 
wavefunctional, thereafter it is a matter of computational effort to obtain 
results.  In comparison, the \DS equations can be derived completely (in the 
sense that, in principle, all loop terms can be written down) but must 
subsequently be truncated in order to furnish useful equations.  Having made 
the connection between the two approaches at leading order here, it seems 
promising that beyond leading order such comparison may provide useful 
insights into both approaches.

The heavy quark limit of the \DS equations in Coulomb gauge also seems a 
very promising avenue for further study.  From the lattice, it is known that 
the physical, Wilson string tension is not simply the coefficient of the 
infrared singularity in the instantaneous temporal gluon propagator (see, 
for example, Ref.~\cite{Iritani:2011zg}).  The Coulomb string tension is 
analytically known to be larger than the Wilson string tension, encapsulated 
in the statement: ``No confinement without Coulomb confinement." 
\cite{Zwanziger:2002sh}.  Combining the heavy quark limit of the \DS 
equations beyond the leading order truncation presented here with the 
Bethe-Salpeter equation, one may hope to study the quark-antiquark potential 
quantitatively.  Thereafter, phenomenological application to the hadron 
spectrum would be a realistic proposition.

\begin{acknowledgments}
The authors gratefully acknowledge useful discussions with G.~Burgio and 
M.~Pak.  This work has been supported by the Deutsche Forschungsgemeinschaft 
(DFG) under contracts no. DFG-Re856/6-2,3.
\end{acknowledgments}



\begin{thebibliography}{99}

\bibitem{Gribov:1977wm}
  V.~N.~Gribov,
  Nucl.\ Phys.\  {\bf B139 } (1978)  1.

\bibitem{Zwanziger:1995cv}
  D.~Zwanziger,
  Nucl.\ Phys.\  {\bf B485}, 185-240 (1997).
  [hep-th/9603203].

\bibitem{Zwanziger:1998ez}
  D.~Zwanziger,
  Nucl.\ Phys.\  {\bf B518 } (1998)  237-272.
  
\bibitem{Szczepaniak:1995cw}
  A.~Szczepaniak, E.~S.~Swanson, C.~-R.~Ji, S.~R.~Cotanch,
  Phys.\ Rev.\ Lett.\  {\bf 76}, 2011-2014 (1996).
  [hep-ph/9511422].

\bibitem{LlanesEstrada:2000jw}
  F.~J.~Llanes-Estrada, S.~R.~Cotanch, P.~J.~de A. Bicudo, 
J.~E.~F.~T.~Ribeiro, A.~P.~Szczepaniak,
  Nucl.\ Phys.\  {\bf A710}, 45-54 (2002).
  [hep-ph/0008212].

\bibitem{Szczepaniak:2001rg}
  A.~P.~Szczepaniak, E.~S.~Swanson,
  Phys.\ Rev.\  {\bf D65 } (2002)  025012.
  [hep-ph/0107078].

\bibitem{Szczepaniak:2003ve}
  A.~P.~Szczepaniak,
  Phys.\ Rev.\  {\bf D69 } (2004)  074031.
  [hep-ph/0306030].

\bibitem{Feuchter:2004mk}
  C.~Feuchter, H.~Reinhardt,
  Phys.\ Rev.\  {\bf D70 } (2004)  105021.
  [hep-th/0408236].

\bibitem{Reinhardt:2004mm}
  H.~Reinhardt, C.~Feuchter,
  Phys.\ Rev.\  {\bf D71 } (2005)  105002.
  [hep-th/0408237].

\bibitem{Campagnari:2010wc}
  D.~R.~Campagnari, H.~Reinhardt,
  Phys.\ Rev.\  {\bf D82 } (2010)  105021.
  [arXiv:1009.4599 [hep-th]].

\bibitem{Pak:2011wu}
  M.~Pak, H.~Reinhardt,
    [arXiv:1107.5263 [hep-ph]].

\bibitem{Adler:1984ri}
  S.~L.~Adler, A.~C.~Davis,
  Nucl.\ Phys.\  {\bf B244 } (1984)  469.

\bibitem{Schutte:1985sd}
  D.~Schutte,
  Phys.\ Rev.\  {\bf D31 } (1985)  810-821.
  
\bibitem{Christ:1980ku}
  N.~H.~Christ and T.~D.~Lee,
  Phys.\ Rev.\  D {\bf 22}, 939 (1980)
  [Phys.\ Scripta {\bf 23}, 970 (1981)].

\bibitem{Watson:2006yq}
  P.~Watson, H.~Reinhardt,
  Phys.\ Rev.\  {\bf D75 } (2007)  045021.
  [hep-th/0612114].

\bibitem{Lichtenegger:2009dw}
  K.~Lichtenegger, D.~Zwanziger,
    [arXiv:0911.5435 [hep-ph]].

\bibitem{Alkofer:2009dm}
  R.~Alkofer, A.~Maas, D.~Zwanziger,
  Few Body Syst.\  {\bf 47}, 73-90 (2010).
  [arXiv:0905.4594 [hep-ph]].

\bibitem{Watson:2007mz}
  P.~Watson, H.~Reinhardt,
  Phys.\ Rev.\  {\bf D76 } (2007)  125016.
  [arXiv:0709.0140 [hep-th]].

\bibitem{Watson:2007vc}
  P.~Watson, H.~Reinhardt,
  Phys.\ Rev.\  {\bf D77 } (2008)  025030.
  [arXiv:0709.3963 [hep-th]].

\bibitem{Popovici:2008ty}
  C.~Popovici, P.~Watson, H.~Reinhardt,
  Phys.\ Rev.\  {\bf D79 } (2009)  045006.
  [arXiv:0810.4887 [hep-th]].

\bibitem{Watson:2008fb}
  P.~Watson, H.~Reinhardt,
  Eur.\ Phys.\ J.\  {\bf C65 } (2010)  567-585.
  [arXiv:0812.1989 [hep-th]].

\bibitem{Reinhardt:2008pr}
  H.~Reinhardt and P.~Watson,
  Phys.\ Rev.\  D {\bf 79}, 045013 (2009)
  [arXiv:0808.2436 [hep-th]].

\bibitem{Popovici:2010mb}
  C.~Popovici, P.~Watson, H.~Reinhardt,
  Phys.\ Rev.\  {\bf D81 } (2010)  105011.
  [arXiv:1003.3863 [hep-th]].

\bibitem{Popovici:2010ph}
  C.~Popovici, P.~Watson, H.~Reinhardt,
  Phys.\ Rev.\  {\bf D83 } (2011)  025013.
  [arXiv:1010.4254 [hep-ph]].

\bibitem{Popovici:2011yz}
  C.~Popovici, P.~Watson, H.~Reinhardt,
  Phys.\ Rev.\  {\bf D83 } (2011)  125018.
  [arXiv:1103.4786 [hep-ph]].

\bibitem{Cucchieri:2000hv}
  A.~Cucchieri, D.~Zwanziger,
  Phys.\ Rev.\  {\bf D65 } (2001)  014002.
  [hep-th/0008248].

\bibitem{Cucchieri:2000gu}
  A.~Cucchieri, D.~Zwanziger,
  Phys.\ Rev.\  {\bf D65 } (2001)  014001.
  [hep-lat/0008026].

\bibitem{Langfeld:2004qs}
  K.~Langfeld, L.~Moyaerts,
  Phys.\ Rev.\  {\bf D70 } (2004)  074507.
  [hep-lat/0406024].
  
\bibitem{Cucchieri:2007uj}
  A.~Cucchieri, A.~Maas, T.~Mendes,
  Mod.\ Phys.\ Lett.\  {\bf A22 } (2007)  2429-2438.
  [hep-lat/0701011].

\bibitem{Quandt:2008zj}
  M.~Quandt, G.~Burgio, S.~Chimchinda, H.~Reinhardt,
  PoS {\bf CONFINEMENT8 } (2008)  066.
  [arXiv:0812.3842 [hep-th]].

\bibitem{Nakagawa:2009is}
  Y.~Nakagawa, A.~Nakamura, T.~Saito, H.~Toki,
  PoS {\bf LAT2009 } (2009)  230.
  [arXiv:0911.2550 [hep-lat]].

\bibitem{Nakagawa:2011ar}
  Y.~Nakagawa, A.~Nakamura, T.~Saito, H.~Toki,
  Phys.\ Rev.\  {\bf D83 } (2011)  114503.
  [arXiv:1105.6185 [hep-lat]].

\bibitem{Iritani:2011zg}
  T.~Iritani, H.~Suganuma,
  Phys.\ Rev.\  {\bf D83 } (2011)  054502.
  [arXiv:1102.0920 [hep-lat]].

\bibitem{Zwanziger:2002sh}
  D.~Zwanziger,
  Phys.\ Rev.\ Lett.\  {\bf 90}, 102001 (2003).
  [hep-lat/0209105].

\bibitem{Eichten:1980mw}
  E.~Eichten, F.~Feinberg,
  Phys.\ Rev.\  {\bf D23 } (1981)  2724.
  
\bibitem{Zwanziger:1991gz}
  D.~Zwanziger,
  Nucl.\ Phys.\  {\bf B364 } (1991)  127-161.

\bibitem{hep-th/0605115} 
  W.~Schleifenbaum, M.~Leder and H.~Reinhardt,
  Phys.\ Rev.\ D\ {\bf 73}, 125019  (2006)
  [hep-th/0605115].

\bibitem{Watson:2010cn}
  P.~Watson, H.~Reinhardt,
  Phys.\ Rev.\  {\bf D82 } (2010)  125010.
  [arXiv:1007.2583 [hep-th]].

\bibitem{Alkofer:2005ug}
  R.~Alkofer, M.~Kloker, A.~Krassnigg, R.~F.~Wagenbrunn,
  Phys.\ Rev.\ Lett.\  {\bf 96 } (2006)  022001.
  [hep-ph/0510028].  

\end{thebibliography}
\end{document}